\documentclass[a4paper,11pt]{article}
\pdfoutput=1 

\usepackage{jcappub} 
\usepackage{aas_macros}

\usepackage[T1]{fontenc} 

\usepackage{amsmath}
\usepackage{dsfont}
\usepackage[yyyymmdd]{datetime}
\usepackage[export]{adjustbox}
\usepackage{braket}
\usepackage{graphicx}
\usepackage{subcaption}
\usepackage{multirow}
\usepackage{enumitem}
\usepackage{siunitx}
\usepackage{orcidlink}

\usepackage{xcolor}

\newcommand{\cC}{\mathcal{C}}

\newcommand{\cH}{\mathcal{H}}

\newcommand{\cN}{\mathcal{N}}
\newcommand{\cO}{\mathcal{O}}

\newcommand{\cR}{\mathcal{R}}

\renewcommand{\SS}{\mathbb{S}}

\newcommand{\eps}{\varepsilon}

\newcommand{\pp}{\mathrm{p}}

\newcommand{\bs}{\boldsymbol}
\newcommand{\dd}{\mathrm{d}}
\newcommand{\T}{^\top}  

\DeclareSIUnit \pc {pc}
\DeclareSIUnit \h {\ensuremath{\mathit{h}}}

\newcommand{\hurl}[1]{\href{https://#1}{\nolinkurl{#1}}}

\title{\boldmath Benchmarking field-level cosmological inference from galaxy redshift surveys}

\author[a]{Hugo Simon\orcidlink{0009-0004-3707-9282},}
\author[b]{François Lanusse\orcidlink{0000-0001-7956-0542},}
\author[a]{Arnaud de Mattia\orcidlink{0000-0003-0920-2947}}

\affiliation[a]{Université Paris-Saclay, CEA, Irfu,\\91191, Gif-sur-Yvette, France}
\affiliation[b]{Université Paris-Saclay, Université Paris Cité, CEA, CNRS, AIM,\\91191, Gif-sur-Yvette, France}

\emailAdd{hugo.simon@cea.fr}
\emailAdd{francois.lanusse@cnrs.fr}
\emailAdd{arnaud.de-mattia@cea.fr}

\abstract{Field-level inference has emerged as a promising framework to fully harness the cosmological information encoded in next-generation galaxy surveys. It involves performing Bayesian inference to jointly estimate the cosmological parameters and the initial conditions of the cosmic field, directly from the observed galaxy density field. Yet, the scalability and efficiency of sampling algorithms for field-level inference of large-scale surveys remain unclear. To address this, we introduce a standardized benchmark using a fast and differentiable simulator for the galaxy density field based on \texttt{JaxPM}. We evaluate a range of sampling methods, including standard Hamiltonian Monte Carlo (HMC), No-U-Turn Sampler (NUTS) without and within a Gibbs scheme, and both adjusted and unadjusted microcanonical samplers (MAMS and MCLMC). These methods are compared based on their efficiency, in particular the number of model evaluations required per effective posterior sample.
Our findings emphasize the importance of carefully preconditioning latent variables and demonstrate the significant advantage of (unadjusted) MCLMC for scaling to $\geq 10^6$-dimensional problems. We find that MCLMC outperforms adjusted samplers by over an order-of-magnitude, with a mild scaling with the dimension of our inference problem. This benchmark, along with the associated publicly available code, is intended to serve as a basis for the evaluation of future field-level sampling strategies. The code is openly available at \hurl{github.com/hsimonfroy/benchmark-field-level}.
}

\begin{document}
\maketitle
\flushbottom

\section{Introduction}
\label{sec:intro}

In the current cosmological model the evolution of the Universe is governed by a small set of cosmological parameters. Inflation gives rise to an initial field of primordial fluctuations, which evolve under gravity to form the Large Scale Structure (LSS) of the Universe. Galaxy surveys such as DESI~\citep{desi_2024_ValidationProgram}, Euclid~\citep{euclid_2024}, and Vera Rubin's Legacy Survey of Space and Time~\citep{lsst_2009} serve as powerful probes of LSS by mapping the spatial distribution of galaxies, which traces the underlying matter density field and provides a direct observational signature of the evolved primordial fluctuations.

\begin{figure}[htbp]
    \centering
    \includegraphics[width=0.8\linewidth]{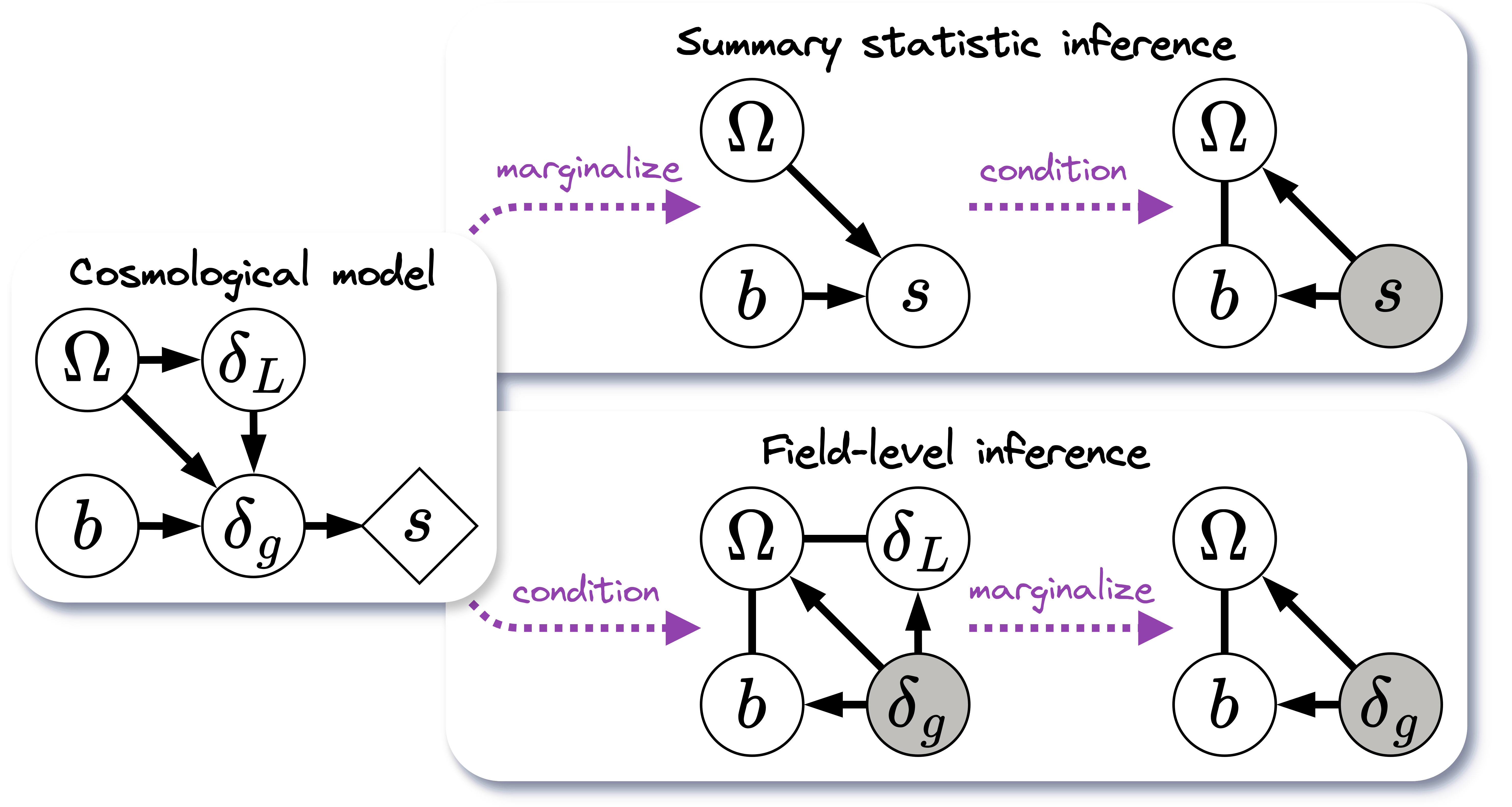}
    \caption{Two approaches to inferring cosmology $\Omega$ from the galaxy density field $\delta_g$, that depends on $\Omega$, initial field $\delta_L$, and nuisance parameters $b$ via a cosmological model (left panel). Top panel: inference based on summary statistics $s(\delta_g)$ marginalizes over $\delta_L$ and $\delta_g$ (e.g. by fitting a Gaussian model or a Neural Density Estimator on simulations) to obtain marginal model $\pp(\Omega)\,\pp(b)\,\pp(s \mid, \Omega, b)$, then conditions on $s$ (e.g. by MCMC sampling) to obtain final posterior $\pp(\Omega, b\mid s)$. Bottom panel: field-level inference conditions on $\delta_g$ (e.g. by MCMC sampling) to obtain full posterior $\pp(\Omega, b, \delta_L \mid \delta_g)$, then marginalizes over $\delta_L$ (by discarding $\delta_L$ samples) to obtain final posterior $\pp(\Omega, b \mid \delta_g)$. In this Bayesian network formalism, white (resp. grey, diamond) nodes denote latent (resp. observed, deterministic) variables, and directed (resp. undirected) edges denote computed (resp. uncomputed) dependencies.}
    \label{fig:dag_comparison}
\end{figure}

A common approach to extracting information from galaxy surveys is to compress the observable galaxy density field $\delta_g$ into a summary statistics $s(\delta_g)$. As illustrated in figure~\ref{fig:dag_comparison} (top panel), a marginal likelihood $s \mid \Omega, b$, is then constructed to relate cosmological parameters $\Omega$ and a reduced set of nuisance parameters $b$ to the summary statistics $s$, effectively marginalizing over the stochasticity of the fields $\delta_L$ and $\delta_g$. This marginalization can be performed with a Gaussian model, e.g. with analytical or simulation-based mean and covariance \citep{desi_2024_FullShapeY1}, or with more general techniques of Neural Density Estimation, e.g. normalizing flows, also fitted on simulations \citep{hahn_2023_SimBIGFirstCosmological}. Finally, the posterior distribution $\Omega, b \mid s$ is inferred, often using Markov Chain Monte Carlo (MCMC) sampling. The choice of summary statistics is guided by a trade-off between analytical or variational tractability of its likelihood, and the amount of cosmological information retained.

A well-motivated choice for $s(\delta_g)$ is the power spectrum $P$ of $\delta_g$. On large scales, where the field remains approximately Gaussian, the power spectrum is close to being sufficient statistics, i.e. $\Omega \mid P \simeq \Omega \mid \delta_g$. Additionally, the power spectrum likelihood is relatively tractable, as it can often be well approximated by a Gaussian distribution with an analytical mean. Yet, gravity-driven structure formation is non-linear, making the galaxy density field non-Gaussian. In this case, the power spectrum is not sufficient, hence the development of analyses relying on alternative statistics \citep{beyond2pt_2024_DataChallenge}, such as higher order statistics (3PCF \citep{sugiyama_2023_3PCFBOSS}, bispectrum \citep{philcox_boss_2022}), object correlations (peaks \citep{desy3_peaks_2022}, voids \citep{Nadathur_2019, hawken_2020_CosmicVoidsBOSS}, density splits \citep{Paillas_2024, pinon_2025_TheoreticalDensitySplit},...), multiscale counts (wavelet scattering transform \citep{valogiannis_2023}, 1D-PDFs \citep{codis_2016}, Minkowski functionals \citep{liu_2023, jiang_2024_MinkowskiFunctionalsModifiedGravity}, k-nearest neighbors \citep{banerjee_2021}, Betti numbers \citep{Ouellette_2025}...), or automatically-learned statistics (CNN compression \citep{hahn_2023_SimBIGFirstCosmological} or GNN compression \citep{VillanuevaDomingo_2022_GNN_sensitivity}). Nonetheless, striking a balance between the interpretability of the compressed statistic, the tractability of its likelihood, and its ability to probe the cosmological signal exhaustively, remains a fundamental challenge.

\begin{figure}[htbp]
    \centering
    \includegraphics[width=\textwidth]{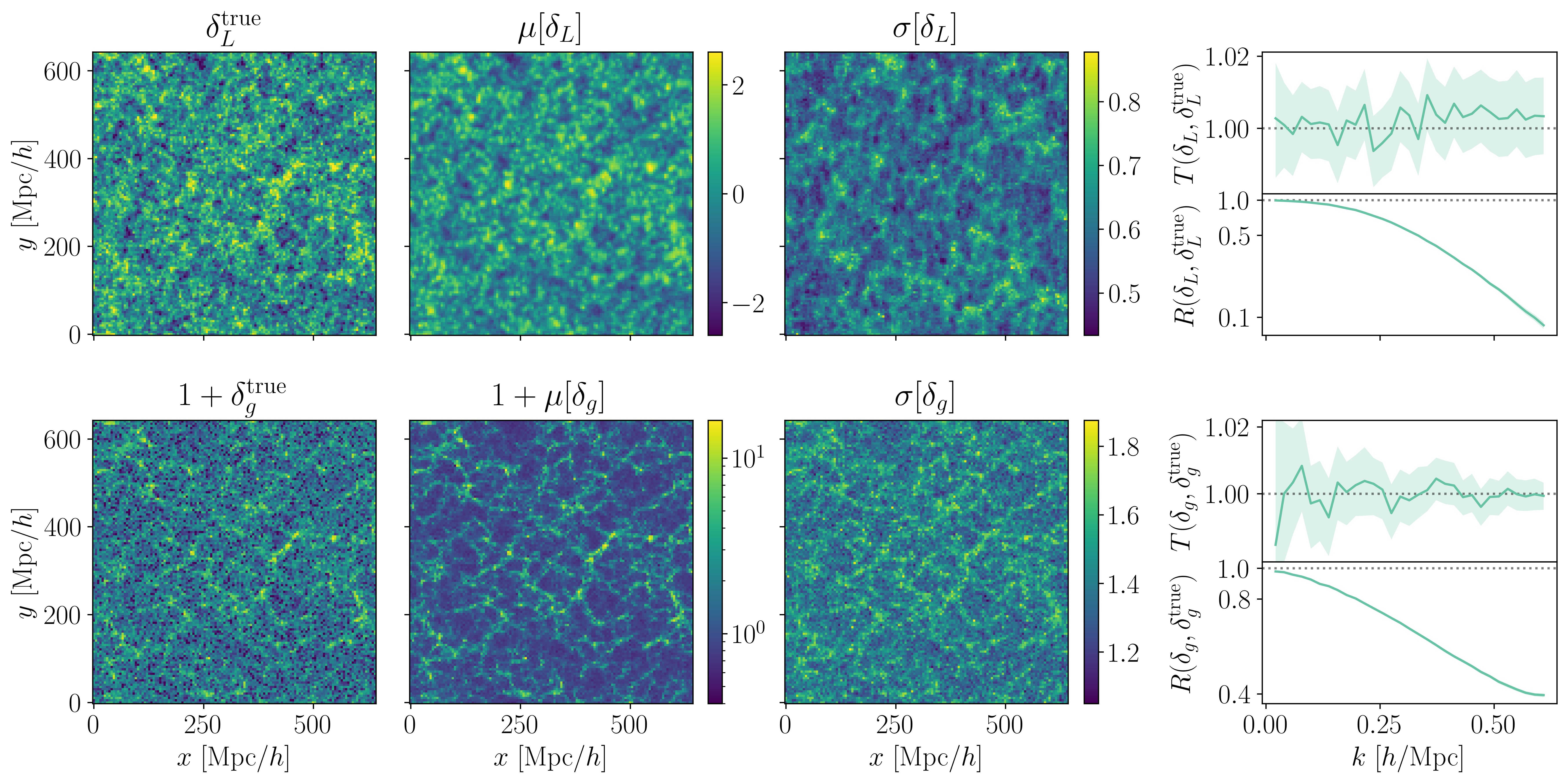} 
    \caption{Posterior of the initial linear matter field $\delta_L$ (top row) and the observable galaxy density field $\delta_g$ (bottom row), obtained by field-level inference from the observed galaxy density field $\delta_g^\mathrm{true}$. Panels from left to right display respectively the voxel true values, posterior means, and posterior standard deviations, for a thin slice of these fields. Rightmost panels show the posterior median and $95\%$-credible intervals of (top) the transfer function $T(\delta_i, \delta_j) = \sqrt{P_{ii} / P_{jj}}$ and (bottom) the spectra correlation $R(\delta_i, \delta_j) = P_{ij} / \sqrt{P_{ii} P_{jj}}$ (with $P_{ij}(k) = \braket{\delta_i(k) \delta_j(k)^*}$) between the sampled field and the true field associated with the observation. Both the linear matter and galaxy density power spectra are well recovered, while the fields themselves are well reconstructed at large scale (spectra correlation close to 1).}
    \label{fig:post_fields}
\end{figure}

An alternative approach, known as field-level inference, bypasses the need for summary statistics by directly conditioning on the observation $\delta_g$. It proceeds so by jointly sampling the initial linear field $\delta_L$ with cosmology $\Omega$ and nuisance parameters $b$, as illustrated in figure~\ref{fig:dag_comparison} (bottom panel). The marginal posterior $\Omega,b \mid \delta_g$ is then readily obtained by discarding $\delta_L$ samples. These samples also enable a posterior reconstruction of the Universe's history at the field level, as illustrated by figure \ref{fig:post_fields} or by~\cite{Lavaux_Jasche_Leclercq_2019}. By construction, this approach extracts all the information encoded in the (observed and discretized) galaxy density field and is therefore quasi lossless. However, applying it to the volume of a large-scale survey like DESI ($V_\mathrm{eff} \approx \qty{20}{(\giga\pc/\h)^3}$), discretized at a typical non-linear scale of \qty{5}{\mega\pc/\h}, results in inferring a latent field with $\operatorname{dim}(\delta_L) \approx 10^9$ parameters. Field-level inference thus shifts the challenge from performing analytical or variational marginalization to solving a high-dimensional sampling problem.

To tackle this, state-of-the-art sampling methods, such as the Hamiltonian Monte Carlo (HMC) algorithm \citep{duane_1987_HMC, Neal_2011}, leverage the gradient information to efficiently explore high-dimensional posteriors. In practice, the requirement for the gradient is a mild constraint on the model implementation, as modern computational frameworks like \texttt{JAX}\footnote{\hurl{github.com/jax-ml/jax}} \citep{jax2018github}
enable automatic differentiation, eliminating the need for manual derivations. HMC is used in most field-level inference applications \cite{Jasche_2013_borg_intro, Lavaux_Jasche_Leclercq_2019, Jasche_2019_borg_vorticity, Porqueres_2020_borg_lyman, boruah_2022_FieldLevelConvergence, Porqueres_2023_borg_wl, nguyen_2024_InformationFieldLevel}. The No U-Turn Sampler (NUTS) \citep{Hoffman_Gelman_2014_NUTS}, a variation of the standard HMC algorithm with automatically tuned trajectory length, has been used by \cite{Zeghal_2024, horowitz_2025_DiffHydro}. \cite{Bayer_Seljak_Modi_2023} recently proposed a different sampler dubbed MicroCanonical Langevin Monte Carlo (MCLMC) to perform cosmological field-level inference, showing it can require over an order-of-magnitude less model evaluations than HMC. Moreover, the choice of reparametrization, i.e. model preconditioning strategies, generally affects the efficiency of these sampling methods. A comprehensive comparison of these strategies is required to identify the most efficient and robust approaches, to assess their scalability to next-generation galaxy surveys, and to guide future algorithmic developments.

In this work, we conduct a comprehensive benchmark of MCMC samplers and preconditionings, focusing on field-level modeling and inference for galaxy clustering data. Our goal is to identify strategies that enable reliable posterior sampling while minimizing the number of computationally expensive simulations. Additionally, we investigate how modeling choices, such as resolution, volume, and complexity in large-scale structure formation modeling impact the high-dimensional sampling process.

The structure of the paper is as follows. In section~\ref{sec:methodo} we present the model, samplers, conditioning strategies and metrics for the benchmark. Comparison of the samplers, conditioning strategies, and impact of modeling complexity are presented and discussed in section~\ref{sec:expe}. We conclude in section~\ref{sec:conclusion}.

\section{Methodology}
\label{sec:methodo}
\subsection{Model}
\label{sec:model}
To benchmark various sampling strategies under the most relevant conditions, we implement a forward model that encapsulates the key physical processes that could impact high-dimensional field-level sampling. These include the initial matter field generation, its non-linear gravitational evolution, the galaxy-matter connection, redshift-space distortions (RSD), and observational noise. The analysis is performed in a self-specified context, where the observations are generated from the same model used for inference. This idealized setting allows us to isolate and study the performance of sampling algorithms without the confounding effects of model misspecification \citep{Nguyen_2021_impact_model, kostic_2023_ConsistencyFieldLevelEFT}. Questions related to the convergence of the density and velocity fields at the probed scales, as well as survey-specific modeling choices, such as observational systematics, are deferred to future work. By focusing on the fundamental components of galaxy clustering models while remaining agnostic to specific survey characteristics, this modeling is designed to be broadly informative for upcoming applications of field-level inference to galaxy surveys.

In practice, we make use of \texttt{JAX}, a \texttt{Python} library providing GPU acceleration, Just-In-Time (JIT) compilation, automatic vectorization/parallelization, and differentiation. We implement our model via \texttt{NumPyro}\footnote{\hurl{github.com/pyro-ppl/numpyro}} \citep{numpyro_2019}, a Probabilistic Programming Language (PPL) powered by \texttt{JAX}, which facilitates model definition, differentiation, and sampling. Our hierarchical model is thus highly modular, and contains the following stages:

\begin{enumerate}[label=\textbf{\arabic*}.]

\item \textbf{Initial matter field:} cosmological parameters $\Omega$ include the matter density $\Omega_\mathrm{m}$, and the amplitude of linear fluctuations (smoothed at a scale of $\qty{8}{\mega\pc/\h}$) $\sigma_8$. The linear matter field $\delta_L$ is then sampled as a Gaussian field, with power spectrum determined by these parameters via the fitting formula of~\cite{Eisenstein_1998} implemented in \texttt{jax-cosmo}\footnote{\hurl{github.com/DifferentiableUniverseInitiative/jax_cosmo}} \citep{campagne_2023_jaxcosmo}.

\item \textbf{LSS formation:} matter particles are initialized on a regular grid and displaced following first/second order Lagrangian Perturbation Theory (1LPT/2LPT), or through a Particle Mesh (PM) N-body scheme. We rely on \texttt{JaxPM}\footnote{\hurl{github.com/DifferentiableUniverseInitiative/JaxPM}}, which is a \texttt{JAX} version of \texttt{FlowPM} \citep{Modi_2021_flowpm_intro}. To compute N-body displacement in particular, we use 5 steps of the BullFrog solver, recently proposed by \cite{rampf_2024_BullFrogMultistepPerturbation}, and implemented within \texttt{JaxPM} framework. In addition to faster convergence, BullFrog does not require prior LPT initialization, eliminating the sensitivity to the starting redshift of other PM solvers, like FastPM \citep{Feng_2016_fastpm_intro}.

\item \textbf{Redshift-Space Distortions:} particles are advected once more according to their velocities $\dot{\bs q}$ projected along line-of-sight to account for contribution of galaxy peculiar velocity to redshift measurement. The RSD displacement depends on the Hubble function $H$,
evaluated at the simulation redshift $z$, and is given by
\begin{equation}
    \Delta \bs q = H(z)^{-1} (\dot{\bs q} \cdot \hat{\bs \eta})\ \hat{\bs \eta}
\end{equation}
where line-of-sight $\hat{\bs \eta}$ is fixed and taken to be one axis of the grid.

\item \textbf{Galaxy bias:} the galaxy density field is linked to the matter field via a second order Lagrangian bias expansion \citep{matsubara_2008_lagrangian, Modi_Chen_White_2020}. Formally, a particle at position $\bs q$ and observed at redshift $z$ is assigned the weight
\begin{equation}
    w_g(\bs q, z) := (1 + b_1 \delta_L + b_2 (\delta_L^2 - \braket{\delta_L^2}) + b_{s^2} (s^2 - \braket{s^2}) + b_{\nabla^2} \nabla^2\delta_L)(\bs q, z)
\end{equation}
where $\delta_L(\bs q, z) = D(z) \delta_L(\bs q)$ with $D$ the linear growth factor and $s^2 = s^{ij} s_{ij}$ with $s_{ij} = [\partial_i \partial_j \nabla^{-2} - \delta_{ij}^{K}/3] \delta_L$ the tidal tensor.
\item \textbf{Observational noise:} the biased field is then converted into the observable galaxy density field $\delta_g$, by adding Gaussian noise with variance $\bar{N}_g^{-1}$ to each voxel, where $\bar{N}_g$ is the average number of galaxies per voxel.

\end{enumerate}

In total, the latent parameters include 2 cosmological parameters \(\Omega := (\Omega_\mathrm{m}, \sigma_8)\); 4 bias parameters from the second order Lagrangian bias expansion \(b := (b_1,b_2,b_{s^2},b_{\nabla^2})\); and the linear matter field $\delta_L$, whose dimensionality depends on discretization of the simulated volume. These latent parameters are preconditioned to improve sampling efficiency, which means that the base variables $(\Omega, b, \delta_L)$ are reparametrized into sample variables $(\tilde \Omega, \tilde b, \tilde \delta_L)$ whose posteriors are closer to a standard Gaussian. We refer to section~\ref{sec:intro_conds} for details on the different conditioning strategies. The true values of $\Omega$ and $b$ used in experiments (to generate synthetic data), their priors, as well as the fiducial values taken to compute the initialization and the static preconditioning (see section~\ref{sec:intro_conds}) are referenced in table~\ref{tab:params}. The prior ranges for bias parameters are typical of values measured from N-body simulations populated with Halo Occupation Distribution scheme \citep{desi_2025_HODInformedPriors} and we ensure in the following that computed posteriors are likelihood-dominated.

\begin{table}[htbp]
    \centering
    \begin{tabular}{lllll}
        \hline
        Name & Symbol & Truth & Prior & Fiducial \\
        \hline
        Matter density & $\Omega_\mathrm{m}$ & 0.3 & $\cN(0.3111, 0.5^2, [0.05,1])$ & 0.3111 \\
        Amplitude of linear fluctuations & $\sigma_8$ & 0.8 & $\cN(0.8102, 0.5^2, \left[0, +\infty\right[)$ & 0.8102 \\
        \multirow{4}{*}{Lagrangian galaxy bias} & $b_1$ & 1 & $\cN(1, 0.5^2)$ & 1 \\
        & $b_2$ & 0 & $\cN(0, 2^2)$ & 0 \\
        & $b_{s^2}$ & 0 & $\cN(0, 2^2)$ & 0 \\
        & $b_{\nabla^2}$ & 0 & $\cN(0, 2^2)$ & 0 \\
        {Linear matter power spectrum} & $P_L$ &  \multicolumn{3}{l}{Given by cosmology $(\Omega_\mathrm{m}, \sigma_8)$ } \\
        {Initial linear matter field} & $\delta_L$ & \multicolumn{3}{l}{$\cN(\bs 0, P_L V_\mathrm{cell}^{-1}$)} \\
        {Observable galaxy density field} & $\delta_g$ & \multicolumn{3}{l}{Evolved from initial field $\delta_L$} \\
        \hline
    \end{tabular}
    \caption{Summary of model cosmological and bias parameters, and fields, with their true values used to generate the observations, their priors, and their fiducial values used for inference initialization. $\cN(\bs \mu, C, A)$ denotes a Gaussian variable of mean $\bs \mu$ and covariance $C$, truncated to the set $A$.}
    \label{tab:params}
\end{table}

As a baseline in this work, we simulate a cosmological volume of $V_\mathrm{sim} = (\qty{320}{\mega\pc/\h})^3$ discretized into a $64^3$ mesh, resulting in a voxel resolution of \qty{5}{\mega\pc/\h}. Also, $\bar{N}_g := \bar{n}_g V_\mathrm{cell}$ is fixed assuming a DESI-like mean galaxy density of $\bar{n}_g = \qty{e-3}{(\mega\pc/\h)^{-3}}$.

\subsection{MCMC samplers}
\label{sec:intro_samplers}

To perform the high-dimensional sampling required in field-level inference, we cannot afford to employ a classical Metropolis-Hastings (MH) algorithm. This algorithm samples a target density via a simple Markov chain where each step is agnostic to local information about the density. This leads to two major issues: (i) as the dimensionality increases, more directions lead away from high-density regions, so that the step size must scale as $\cO(d^{-1/2})$ to maintain a given acceptance rate; (ii) once the dynamics has reached high-density regions, motion becomes locally Brownian (a random walk), meaning $n^2$ steps of size $\eps$ are required to move by $\cO(n \eps)$, which is an inherently slow way to explore the sampling space. 

Addressing these issues requires integrating local information about the target density into the dynamics. Moreover, significant advances in numerical sampling and optimization have been achieved by exploiting their connection to statistical physics: many randomized iterative methods exhibit particle-like dynamics, which can be analyzed statistically and leveraged to inform new developments. The following subsections provide an overview of canonical and microcanonical Langevin MCMC methods to introduce the different samplers that we consider in our benchmark.

\subsubsection{Canonical sampling with Langevin dynamics}
Our goal is to sample from the $d$-variate target density $\pp \propto e^{-U}$. In the potential $U$ evolves a particle of position $\bs q$, momentum $\bs p$, and mass (matrix) $M$. The Hamiltonian of this particle is then given by:

\begin{equation}
    \mathcal H(\bs q, \bs p) = U(\bs q) + \frac 1 2 \bs p^\top M^{-1} \bs p 
\end{equation}

As a result, its motion follows Hamilton equations
\begin{equation}
    \begin{cases}	  
    \dot {{\bs q}} = \partial_{\bs p}\cH =  M^{-1}{{\bs p}}\\	  
    \dot {{\bs p}} = - \partial_{\bs q}\cH =  - \nabla U(\bs q)  \end{cases}
\label{eqn:canonical_hamilt}
\end{equation}
which constrain the particle to the constant-energy surface $\{\bs q, \bs p \mid \cH(\bs q, \bs p) = E\}$, where $E:=\cH(\bs q_0, \bs p_0)$ denotes the initial particle energy. Confined to this surface, the particle cannot sample from a general target density $\pp$ without energy exchange.

Energy exchange can be provided by the particle's surrounding medium, a heat bath, where it collides with other particles at a frequency $\gamma$. The resulting dynamics in the continuous limit of interaction is described by the Langevin equations
\begin{equation}
    \begin{cases}	  \dot {\bs q} = M^{-1}{\bs p}\\	  \dot {\bs p} = - \nabla U(\bs q) - \gamma \bs p + (2 \gamma M)^{1/2} \dot{\bs w}  \end{cases}
\label{eqn:canonical_langevin}
\end{equation}
where $\bs w$ is a Wiener process (so $\dot{\bs w}$ is a Gaussian white noise). These equations add to the Hamiltonian dynamics a dissipation term $- \gamma \bs p$ and a fluctuation term $(2 \gamma M)^{1/2} \dot{\bs w}$, which account respectively for the particle dissipating coherent kinetic energy as heat and gaining dispersed kinetic energy from thermal fluctuations.

Energy exchange allows the particle system to reach thermal equilibrium, where the distribution of particle states is given by the {canonical ensemble} $\pp_\text{C}(\bs q, \bs p) \propto e^{-\mathcal H(\bs q, \bs p)}$. In particular and as desired, the particle position $\bs q$ is distributed as the target $\pp$:

\begin{equation}
    \int \pp_\text{C}(\bs q, \bs p) \dd \bs p = \int \pp(\bs q)\, \cN(\bs p \mid \bs 0, M) \, \dd \bs p = \pp(\bs q)\,.
\end{equation}

\subsubsection{Microcanonical sampling with isokinetic Langevin dynamics}

A limitation of introducing energy exchange into the dynamics is that the target distribution is only recovered by averaging particle trajectories across multiple energy levels, potentially slowing convergence. Hopefully, energy exchange is not necessary if, instead of a classical particle, we consider an isokinetic particle, i.e. a particle of constant kinetic energy $\dot{\bs q}\T M \dot{\bs q} / 2 = |\bs u|^2 / 2 = 1/2$, where $\bs u:= M^{1/2} \dot{\bs q}$ denotes an auxiliary momentum. 

Physically, this isokineticity can be enforced by coupling the particle to a Gaussian IsoKinetic (GIK) thermostat \citep{morriss_1998_ThermostatsAnalysisApplication}, which constantly extracts or provides energy to the particle so that its temperature, as a kinetic energy, is a constant of motion (see appendix~\ref{sec:gik_hamilt}). When placed in the rescaled potential $U/(d-1)$ (recalling $d\geq 2$ is dimension), such particle admits the GIK Hamiltonian\footnote{In contrast to \cite{steeg_2021_HamiltonianDynamicsNonNewtonian, robnik_2022_MicrocanonicalHamiltonianMonte, robnik_2024_FluctuationDissipationMicrocanonical} we adopt the GIK Hamiltonian formulation that does not require any momentum-dependent time rescaling. This preserves $\pp$ as a stationary density of the dynamics, and therefore reconciles the microcanonical formulation from \citep{steeg_2021_HamiltonianDynamicsNonNewtonian, robnik_2022_MicrocanonicalHamiltonianMonte} and the isokinetic formulation from \citep{robnik_2024_FluctuationDissipationMicrocanonical}. In appendix \ref{sec:gik_hamilt}, we show that we can rederive from this Hamiltonian the MCLMC equations of motion.}
\begin{equation}
    \cH(\bs q, \bs p) = \frac {\bs p\T M^{-1} \bs p} {2 m(\bs q)} - \frac{m(\bs q)}{2}
\label{eqn:gik_hamilt}
\end{equation}
where $m := e^{-U/(d-1)}$ and $\bs p:= m M^{1/2} \bs u = m M \dot{\bs q}$ is the canonical momentum. Remarkably, this Hamiltonian's {microcanonical ensemble} $\pp_\text{MC}(\bs q, \bs p) \propto \delta(\cH(\bs q,\bs p))$ coincides exactly with its {isokinetic ensemble} $\pp_\text{IK}(\bs q, \bs u) \propto  e^{-U(\bs q)} \delta(|\bs u|^2 - 1)$. This means that under ergodicity, the particle is sampling these ensembles, and therefore has position $\bs q$ distributed as target $\pp$:
\begin{equation}
\int \pp_\text{MC}(\bs q, \bs p)\, \dd \bs p = \int \pp_\text{IK}(\bs q, \bs u)\, \dd \bs u \propto \int \pp(\bs q)\, \delta(|\bs u|^2 - 1)\, \dd \bs u \propto \pp (\bs q)\,.
\end{equation}


However, merely following the induced Hamiltonian dynamics does not ensure the ergodicity required for sampling the microcanonical/isokinetic ensemble. In fact, if the system exhibits additional symmetries beyond time translation (which already enforces energy conservation), Noether's theorem guarantees the opposite: the particle remains confined to a subset of the constant-energy surface defined by an additional conserved quantity.

To ensure ergodicity, \cite{robnik_2022_MicrocanonicalHamiltonianMonte} reintroduces the particle into a heat bath, where it collides with other particles at a frequency $\gamma$. However, due to the GIK thermostat, these interactions always conserve kinetic energy. For numerical efficiency, it is preferred to compute the dynamics for state $(\bs q, \bs u)$, rather than $(\bs q, \bs p)$, which follows in the continuous limit of interaction the isokinetic Langevin equations \citep{robnik_2024_FluctuationDissipationMicrocanonical} (in Stratonovich form)
\begin{equation}
\begin{cases}
\dot{\bs q} = M^{-1/2} \bs u\\
\dot{\bs u} = P(\bs u)( - M^{-1/2} \nabla U(\bs q) / (d-1) + (2 \gamma)^{1/2} \dot{\bs w})
\end{cases}
\label{eqn:micro_langevin} 
\end{equation}
where $P(\bs u) := I - \bs u \bs u\T$ is the orthogonal projection onto $\operatorname{span}(\bs u)^\perp$. 
Therefore thermal fluctuations only impact the momentum direction $\bs u$ while both Hamiltonian $\cH$ and kinetic energy $|\bs u|^2/2$ are conserved.

\subsubsection{Derived samplers}

The Langevin dynamics modeled by equations~\ref{eqn:canonical_langevin} and \ref{eqn:micro_langevin} are the basis of several sampling algorithms, involving different regimes of interaction with medium $\gamma$ and inertia $M$. In practice, the dynamics are solved numerically, often using symplectic integrators like KDK-Verlet/Leapfrog or McLachlan/Minimum-Norm integrator \citep{mclachlan_1995_NumericalIntegrationOrdinary}. Numerical integration comes with discretization errors that can induce a bias, which can be either eliminated via MH adjustement (accept-reject step), or controlled via the integration step size. 

In this benchmark, we consider three canonical and adjusted samplers, and two microcanonical samplers. Our selection is guided by both the existing literature on field-level inference and the availability of the implementations. It is designed to highlight key methodological distinctions, particularly between canonical vs. microcanonical, and adjusted vs. unadjusted methods.

\paragraph{HMC.} 
The celebrated Hamiltonian Monte Carlo algorithm \citep{Neal_2011} simulates a classical particle evolving according to equations~\ref{eqn:canonical_langevin} in a sparse medium ($\gamma = 0$) for a tunable trajectory time length $L$, before colliding with
a heavy particle, which refreshes its momentum: $\bs p \sim \cN(\bs 0, M)$. This arrival state is then used as a proposal for a MH adjustement step, after which the dynamics repeats.

\paragraph{NUTS.}
No-U-Turn Sampler \citep{Hoffman_Gelman_2014_NUTS} is a variant of HMC that automatically and dynamically adapts the trajectory time length $L$. It constructs sets of proposals along the Hamiltonian trajectories, which are stopped whenever they start to reverse direction, thereby avoiding redundant exploration and the need for manual tuning of $L$.

\paragraph{NUTSwG.}
Any MCMC sampler can be implemented within a Gibbs sampling scheme \citep{Geman_Geman_1984}, where MCMC sampling alternates over subsets of parameters. For instance, assume one wants to sample from three blocks of variables $q_1, q_2, q_3$. Then a MCMC within Gibbs scheme consists in sampling alternatively with the given MCMC from $q_1 \mid q_2, q_3$, then $q_2 \mid q_1, q_3$, then $q_3 \mid q_1, q_2$, using the sample of one block to condition the others. It can be seen as a sampling generalization of alternating gradient descent. We consider a NUTS within Gibbs implementation, referred to as NUTSwG in the following.

\paragraph{MAMS.}
Metropolis-Adjusted Microcanonical Sampler \cite{robnik_2025_MetropolisAdjustedMicrocanonical} is a microcanonical counterpart to canonical HMC. It simulates an isokinetic particle evolving according to equations~\ref{eqn:micro_langevin} in a sparse medium ($\gamma = 0$) for a tunable average trajectory time length $L$, before colliding with a heavy particle, which refreshes its momentum: $\bs u \leftarrow \bs z/ \lvert \bs z \rvert$ where $\bs z \sim \cN(\bs 0, I)$. This arrival state is then used as a proposal for a MH adjustement step, after which the dynamics repeats.

\paragraph{MCLMC.} 
MicroCanonical Langevin Monte Carlo \cite{robnik_2022_MicrocanonicalHamiltonianMonte} simulates an isokinetic particle evolving according to equations~\ref{eqn:micro_langevin} in a dense medium. There, the particle undergoes continuous collisions with lightweight particles, causing its momentum to be partially refreshed at each integration step: $\bs u \leftarrow (\bs u + \sqrt{2\gamma} \bs z) / \lvert \bs u + \sqrt{2\gamma} \bs z\rvert$, where $\bs z \sim \cN(\bs 0, I)$,\footnote{This refreshment step replaces the less direct-to-sample Brownian motion on the hypersphere $\SS^{d-1}$ \citep{mijatovic_2020_ExactSphericalBrownian}, corresponding to the diffusion term $\sqrt{2\gamma} P (u) \dot W$ in equations~\ref{eqn:micro_langevin}. Both preserve $\pp$ as a stationary density and provide the desired ergodicity \citep{robnik_2024_FluctuationDissipationMicrocanonical}, and they even closely agree when $\eps \ll L$.} and the tunable momentum decoherence length $L$ is introduced such that $\gamma := (e^{2 \eps / L} - 1)/(2d)$. Crucially, MCLMC is run without MH adjustement step, meaning that every integration state is directly accepted as a sample. Any potential bias introduced by this unadjusted scheme is controlled by the integration step size $\eps$, as discussed below.

\medskip

On the one hand, adjusted algorithms are asymptotically unbiased, but at the cost of significantly reducing the integration step size as dimension increases in order to maintain a reasonable acceptance rate. Formally, \cite{Neal_2011} show that for a target product density, the number of model evaluations needed to reach a new effective sample grows as $d$ for the MH algorithm, as $d^{1/3}$ for adjusted overdamped ($M \ll \gamma$) Langevin dynamics, and as $d^{1/4}$ for adjusted underdamped ($\gamma \ll M$) Langevin dynamics.

On the other hand, unadjusted algorithms have no unbiased guarantees, but the relation between step size and bias has been theoretically studied for both canonical and microcanonical samplers \citep{durmus_2024_AsymptoticBiasInexact, robnik_2024_ControllingAsymptoticBias}. In particular, \cite{robnik_2024_ControllingAsymptoticBias} show, both formally for a target Gaussian and empirically, that the bias can be controlled via $\braket{\Delta E^2} /d$, the Energy Error Variance Per Dimension (EEVPD) across integration steps, which itself depends on the step size in a dimension-independent manner. Therefore, by tuning the step size to keep the EEVPD below some threshold, one ensures that the total sampling error (in terms of Wasserstein distance or Mean Square Error) is dominated by the Monte Carlo error. Since the Effectively, this EEVPD threshold governs the trade-off between the sampling efficiency (the number of model evaluations needed to reach a new effective sample) and the bias, independently of the dimension. This is a really promising result compared to the scaling of adjusted methods. Admittedly, the sampled posterior is rarely an isotropic Gaussian in practice, so we may expect some dependency of the sampling efficiency to dimension. However, as explained in section \ref{sec:intro_conds}, we can reparameterize the model such that the posterior approaches an isotropic Gaussian.

\subsection{Conditioning strategies}
\label{sec:intro_conds}

For any sampling problem, one is free to choose any bijective reparametrization of the base variables. Formally, sampling from a target density in a variable space $\mathcal X$ is analytically equivalent to sampling from the image density in the transformed space $\tilde{\mathcal X}$, and then transforming back the samples to $\mathcal X$. However, these are not algorithmically equivalent. The density in $\mathcal X$ can be challenging to sample (due to degeneracies, discontinuities, boundaries, etc.), but a transformed space $\tilde{\mathcal X}$ can often be defined where the density is smoother and easier to sample. This section focuses on preconditioning the latent space so that the posterior of latent variables approximates an isotropic Gaussian.

\subsubsection{Model preconditioning}
\label{sec:intro_preconds}

With full knowledge of the posterior, finding a suitable reparameterization would be simplified. But in practice, when sampling is precisely the way we access posterior information, we can only rely on partial posterior knowledge. A simple preconditioning is prior-conditioning, where (\textit{a priori}) sample variables are made isotropic Gaussian. This is a relevant choice of conditioning if the likelihood constraining power is weak or aligned with the prior, as the resulting prior-conditioned posterior would also be approximately isotropic Gaussian. However, if the likelihood constraining power is strong and varies significantly across variables relatively to their priors, prior-conditioning can exacerbate degeneracies in the posterior \citep{Papaspiliopoulos_Roberts_Sköld_2007}. 

In the context of field-level inference, model preconditioning is exemplified by the choice of parametrization for the linear field $\delta_L$. The linear field can easily be prior-conditioned in real space by considering a Gaussian white noise, which is Fourier-transformed and multiplied by the square root of the power spectrum. However, the likelihood typically induces correlations between voxels, leading to a posterior that is ill-conditioned. Alternatively, the linear field can also easily be prior-conditioned in Fourier space with a complex Hermitian Gaussian noise, multiplied by the square root of the power spectrum. As a result of the large-scale isometry-invariance and linear growth of $\delta_L$, the posterior wavevector amplitudes are approximately independent and Gaussian. However, since different scales are constrained with varying power, the posterior can still remain poorly conditioned.

In order to make the $\delta_L$ posterior more isotropic, we approximate it assuming a linear model for the observable galaxy density field, as described hereafter. 
In Fourier space, for any wavevector $\bs k$, the (discretized) linear matter field $\delta_L(\bs k) \mid \Omega$ is distributed as a complex-Gaussian $\cC \cN(0, P_L(\bs k)V_\mathrm{cell}^{-1})$, recalling that $\cC\cN(\bs \mu, C) = \cN(\operatorname{Re}(\bs \mu), C/2) + i \cN(\operatorname{Im}(\bs \mu), C/2)$ for any complex vector $\bs \mu$ and covariance matrix $C$.
In the flat-sky approximation, this field is then evolved linearly by a factor $B_K := (b_E + (\hat{\bs k} \cdot \hat{\bs \eta})^2 f) D$ known as the Kaiser boost \citep{kaiser_1987_RSDBoost}, where $D$ and $f$ are the linear growth factor and growth rate at the observed redshift and cosmology, $b_E := 1+b_1$ is the linear Eulerian bias associated to the linear Lagrangian bias $b_1$, and $\hat{\bs k} \cdot \hat{\bs \eta}$ is the cosine of the angle between wavevector $\bs k$ and line-of-sight $\hat{\bs \eta}$. Finally, as in the reference forward model, we include observational noise as a Gaussian distribution of variance $\bar N_g^{-1}$, the mean number of galaxies per voxel.  
Put together and applying Bayes formula for this Gaussian linear model, we have
\begin{equation}
    \underbrace{\cC\cN (\delta_g \mid  B_K \delta_L, \bar N_g^{-1})}_{\text{likelihood}}\;\underbrace{\cC \cN(\delta_L \mid 0, P_L V_\mathrm{cell}^{-1})}_{\text{prior}} = \underbrace{\cC\cN(\delta_L \mid \mu, \sigma^2)}_{\text{posterior}}\; \underbrace{\cC \cN(\delta_g \mid 0, \bar N_g^{-1} + B_K^2 P_L V_\mathrm{cell}^{-1})}_{\text{evidence}}
\label{eqn:kaiser_post}
\end{equation}
with\begin{align*}
\sigma&:= (\bar N_g B_K^2 + P_L^{-1}V_\mathrm{cell})^{-1/2} \\
\mu &:= \sigma^2\bar N_g B_K \delta_g
\end{align*}
meaning that we \textit{a posteriori} have $\delta_L \mid  \delta_g, \Omega, b \sim \cC\cN(\mu, \sigma^2)$. Hence we define the preconditioned linear field:
\begin{equation}
\tilde \delta_L := (\delta_L - \mu)/\sigma \sim \cC\cN( -\mu/\sigma, P_L V_\mathrm{cell}^{-1}/\sigma^2)
\end{equation}
such that its estimated posterior is as desired $\tilde \delta_L \mid \delta_g, \Omega, b \sim \cC\cN(0, I)$. This reparametrization obviously depends on cosmology and bias parameters, which we specify following these two options (that we compare in section~\ref{sec:expe_conds}):
\begin{enumerate}
    \item Static Kaiser conditioning: we assume fixed fiducial values $\Omega^\mathrm{fid}, b_1^\mathrm{fid}$, as given in table~\ref{tab:params}. 
    \item Dynamic Kaiser conditioning: we use the values of $\Omega, b_1$ of the current sample, thereby adapting the conditioning to the local shape of the posterior.\footnote{This is equivalent to an analytical and spatially-varying Riemannian metric, as leveraged in Riemannian Manifold Hamiltonian Monte Carlo (RMHMC) \citep{girolami_2011_RiemannManifoldLangevinHamiltonian}.}
\end{enumerate}

Cosmological and bias parameters $\Omega, b$ are also transformed into $\tilde \Omega, \tilde b$ based on a rough estimate of their posterior standard deviation $\sigma[\Omega, b \mid \delta_g]$, such that $\tilde \Omega, \tilde b \mid \delta_g$ is approximately a standard Gaussian. To estimate $\sigma[\Omega, b \mid \delta_g]$, we observe that it scales as the inverse square root of the galaxy sample volume and extrapolate the constraints from a fast (low-dimensional) inference performed within a smaller volume.

The preconditioning strategy described above leverages approximate knowledge of the posterior to enhance sampling efficiency. However, at smaller scales, the Kaiser model approximation no longer holds. In addition, when considering observational data, the spatially-varying survey selection function will introduce correlations between wavevector amplitudes. We thus provide in appendix \ref{sec:comp_selec} an approximate generalization of the preconditioning for curved-sky and non-trivial selection, as well as a comparison of sampling performance against the periodic box case. More generally, to further improve the model conditioning where analytic prescriptions break down, we can rely on an adaptive mass matrix.

\subsubsection{Mass matrix conditioning}

The MCMC samplers we introduced in section~\ref{sec:intro_samplers} all admit a mass matrix $M$ as a tunable parameter. Formally, a particle of position $\bs q$ and mass $M$ will follow the same dynamics (up to a time and initial momentum rescaling) as a particle of position $\tilde {\bs p} := M^{1/2} \bs q$ and unit mass $I$, i.e. the mass matrix $M$ is equivalent to a linear conditioning of the sample positions. Therefore, a good candidate for the mass matrix is the inverse covariance of the sample variables \citep{Livingstone_Faulkner_Roberts_2017}.

The adopted strategy is to leverage prior knowledge to analytically precondition $\bs q$ into $\tilde{\bs q}$ as described in section~\ref{sec:intro_preconds}, then rely on adaptive methods to set $M = {\tilde\Sigma}^{-1}$, where ${\tilde\Sigma}$ is an online estimate of the covariance of $\tilde {\bs q}$. In practice, this estimation is performed during a warmup phase using Welford's algorithm for numerical stability, and the mass matrix is kept diagonal  for computational efficiency.

\subsection{Metrics}
\label{sec:metrics}

The performance of an estimation procedure $\hat \theta$ can be quantified via its associated risk $\cR[\hat \theta]$, such as the Mean Square Error (MSE). For a given number $n$ of data samples, the efficiency $e_n := \cR[\tilde \theta_n]/\cR[\hat \theta_n]$ compares the risk of a particular estimator $\hat\theta_n$ to the risk of a reference estimator $\tilde \theta_n$. Alternatively, the number of samples $N_\text{eff}$ needed for the reference risk to equate the particular risk, $\cR[\tilde \theta_{N_\text{eff}}] = \cR[\hat \theta_n]$, is commonly referred to as the Effective Sample Size (ESS).

In our case, we obtain from MCMC sampling a stationary and identically distributed, but possibly correlated, sequence of $n$ samples, from which we are interested to estimate some target quantity $\theta$ (e.g. mean, median, covariance, quantile, credible region...). We can then compare the MSE based on the correlated sequence with the MSE that we would have if the samples were independent. The induced ESS denoted $N_\text{eff}[\theta]$ is then understood as ``the number of independent samples that would yield the same statistical power as the MCMC samples for estimating the quantity $\theta$''. Typically, it is the sample size at which the (Markov Chain) Central Limit Theorem operates for this quantity, i.e. $\sqrt{N_\text{eff}[\theta]}(\hat \theta_n - \theta) \simeq \cN(0, \sigma^2[\theta])$. 

It is common and simpler to focus on the estimation of the mean $\mu$, so in the following, the ESS denoted by $N_\text{eff}$ refers to $N_\text{eff}[\mu]$ unless otherwise specified. Formally, for a stationary sequence of size $n$ and autocorrelation $\rho_t$, the ESS is asymptotically 
\begin{equation}
    N_\text{eff} \simeq \frac{n}{\sum_{t=-\infty}^{+\infty}\rho_t} = \frac{n}{1+2\sum_{t=1}^{+\infty}\rho_t}
\end{equation}
and can be estimated based on sample autocorrelation. We see that in the stationary case, the more correlated the samples are, the smaller the ESS becomes. This aligns with the common (though not always true) intuition that highly correlated samples share redundant information, thus reducing their overall effectiveness.

The ESS is a relevant metric for quantifying the quality of a sample sequence, serving as a proxy for estimation error without requiring knowledge of the true mean value. On the other hand, if their true value is known, we can directly compute the error for any quantity of interest, typically the mean $\mu$ and standard deviation $\sigma$. Given $n_c$ chains, we estimate their mean $\hat\mu_c$ and standard deviation $\hat\sigma_c$, and the (empirical and normalized) MSE is obtained as:
\begin{equation}
    \operatorname{{MSE}}[\theta] := \frac{1}{{n_c}} \sum_{c}\eps_{c}^2[\theta] \quad\quad \text{where in particular}\;  \eps_c[\theta] :=  \frac{\hat \theta_c - \theta}{\sigma \sqrt{n_c}} \times \begin{cases} 1 &\text{if } \theta \text{ is } \mu\\ 
    \sqrt{2}  &\text{if } \theta \text{ is }\sigma\end{cases}\,.
\end{equation}
Normalization of errors ensures that for an asymptotically unbiased estimator $\operatorname{{MSE}}[\theta] \simeq {N_\text{eff}[\theta]}^{-1}{\chi^2(n_c)}/{n_c}$, making it commensurable with the inverse ESS.

Given the high dimensionality of our problem, it is often necessary to aggregate metrics across sets of variables, such as cosmological, bias, and linear field parameters. The multivariate generalization of MSE implies that we respectively take the arithmetic mean of empirical MSEs and the harmonic mean of ESSs.

Model evaluation (especially its gradient) is our main computational bottleneck for cosmological inference. We therefore characterize the efficiency of a MCMC sampler by the number of model evaluations required to yield one effective sample $N_\text{eval} / N_\text{eff}[\theta]$, and commensurably, to yield a unit error $N_\text{eval} \times \operatorname{MSE}[\theta]$. These metrics are expected to be relatively stable over evaluations, so we check their convergence before reporting them.

Finally, unadjusted samplers may introduce an asymptotic bias in the sampled posterior (see appendix~\ref{sec:bias_mclmc}), that is not directly quantified by the ESS or the MSE individually\footnote{A high MSE can be caused by both an inefficient sampling (low ESS) or a bias.}. However, since in the unbiased case, $\operatorname{{MSE}}[\mu]$ is of the order of the inverse ESS, finding $\operatorname{{MSE}}[\mu] \gg N_\text{eff}^{-1}$ would reveal a non-negligible bias.

\section{Numerical Results}
\label{sec:expe}

In this section we benchmark a comprehensive range of sampling methods and conditioning strategies according to the metrics specified in section~\ref{sec:metrics} for the model presented in section~\ref{sec:model}. The task is to sample from the posterior $\Omega, b, \delta_L \mid \delta_g^\mathrm{true}$ given an observation $\delta_g^\mathrm{true}$ generated from the same model. In sections~\ref{sec:expe_samplers} and~\ref{sec:expe_conds}, the forward model performs $\text{1LPT}$ displacement in a $V_\mathrm{sim} =(\qty{320}{\mega\pc/\h})^3$ volume discretized into a $64^3$ mesh (voxel resolution of $\qty{5}{\mega\pc/\h}$). In section~\ref{sec:expe_scaling}, we vary these settings.

We run for each inference task 8 parallel chains, on $8\times 80$GB NVIDIA A100 GPU node. Our sampling pipeline is built upon \texttt{JAX}-powered packages \texttt{NumPyro} and \texttt{BlackJAX}\footnote{\hurl{blackjax-devs.github.io/blackjax}} \citep{blackjax_2024}.

\subsection{Initialization}

\begin{figure}[htbp]
    \centering
    \includegraphics[width=0.9\linewidth]{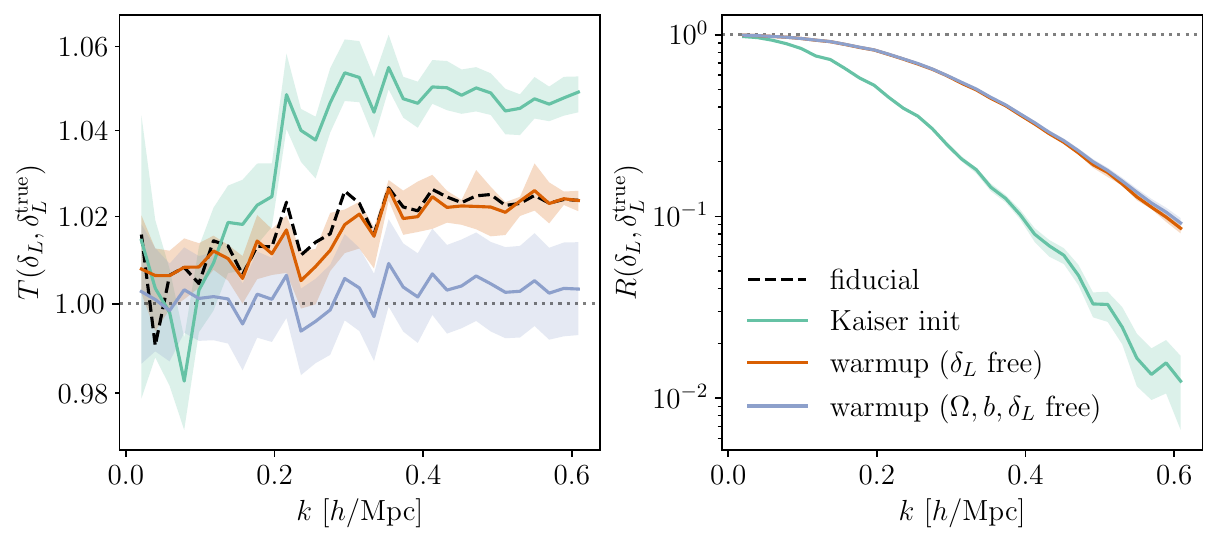}
    \caption{State of the chains at the initialization and at the end of the warmup phases, showed via the median and the $95\%$-credible region of the transfer function (left) and spectra correlation (right) between the sampled linear field and the true linear field (as defined in figure \ref{fig:post_fields}). The initialization based on the Kaiser posterior approximation (turquoise) provides a relevant reconstruction of the linear field $\delta_L$ at large scales (spectra correlation close to 1). From there, the subsequent warmup phases achieve convergence, first (orange) towards a fiducial power spectrum (assuming a fixed fiducial cosmology $\Omega^\mathrm{fid}$ and bias parameters $b^\mathrm{fid}$), then (indigo) towards the true power spectrum (jointly inferring $\Omega, b$ and $\delta_L$).}
    \label{fig:init}
\end{figure}

Simply drawing $\delta_L$ samples from the prior would produce an initial field with the fiducial power spectrum, but its amplitudes and phases would be distributed between the different wavevectors regardless of the observable field $\delta_g$. Instead, we seed the chains with samples from the Kaiser posterior, given by equations~\ref{eqn:kaiser_post} and fiducial parameters. This approximation results in an inaccurate initial linear power spectrum but aligns amplitudes and phases closer to their high-density regions. This is especially the case for the largest scales as reflected by the spectra correlation\footnote{Because spectra correlation is complex-valued, we only show its norm, though we may miss some potential phase shifts.} on figure~\ref{fig:init}. We observe that this initialization strategy indeed accelerates convergence.

We then perform a first warmup phase where exclusively $\delta_L$ is inferred while maintaining other (cosmological and bias) parameters fixed at their fiducial values. This allows the high-dimensional $\delta_L$ to converge towards its high-density region while preventing its dynamics to drag the other parameters away. This warmup phase is assumed successful when the sample power spectrum fluctuates around the fiducial one, as shown by the transfer function on figure~\ref{fig:init}. Note that if we had initialized the linear field from the start with the fiducial power spectrum, we would not be able to rapidly distinguish the chain convergence from the chain being stuck.

Next, a second warmup phase is performed where all parameters are jointly sampled. This is also the phase where sampler hyperparameters are tuned, especially step size and mass matrix. This second warmup is assumed successful when the sampler hyperparameters of the different chains are in qualitative agreement. 

Finally, the chains are run until reaching at least 500 ESS and a Gelman-Rubin statistic of at most $1.01$ for each parameter (which is roughly equivalent, see \cite{vats_knudson_2021_RevisitingGelmanRubin}).

In total, about one fifth of the total number model evaluations is spent for warmup (to be conservative), and we do not include these evaluations in the benchmark results. We observe that a short warmup time is, as expected, correlated with a high sampling efficiency in the stationary regime.

\subsection{MCMC samplers}
\label{sec:expe_samplers}

We start by comparing the performance of the different samplers mentioned in section~\ref{sec:intro_samplers}. For all samplers, we use the Kaiser dynamic preconditioning and tune a (diagonal) mass matrix. Their specific configurations are described hereafter:

\paragraph{HMC, NUTS, NUTSwG.} We solve the equations of motion~\ref{eqn:canonical_hamilt} with the KDK-Verlet integrator and tune the step size $\eps$ to target a $65\%$ acceptance rate, theoretically optimal for isotropic Gaussian densities \cite{Neal_2011}, which we corroborated in our tests.\footnote{The relevant range of the target acceptance rate for HMC-like algorithms is usually $[0.6, 0.9]$, with higher values better suited to more complex densities \citep{Hoffman_Gelman_2014_NUTS, betancourt_2014_OptimizingStepSizeHMC}.}

\paragraph{HMC.} The trajectory time length $L$ is based on the average trajectory length of NUTS, following the heuristic $L_\text{HMC} = \braket{L_\text{NUTS}} / 2$.\footnote{Hamiltonian trajectory length is fixed here, although randomizing it could improve certain performance metrics in some cases \citep{jiang_2023_DissipationIdealHamiltonian}.}

\paragraph{NUTSwG.} We split the sampled variables into two blocks: (i) the linear field $\delta_L$ and (ii) cosmological and bias parameters $(\Omega, b)$, in a similar fashion to~\citep{Jasche_2019_borg_vorticity, Porqueres_2023_borg_wl}\footnote{This choice can be partly motivated by a lack of differentiable computation of cosmological quantities. In the referenced works, each cosmological and bias parameter is itself treated in a separate block.}. A NUTS within Gibbs scheme for these blocks means that we alternatively sample with NUTS from $\delta_L \mid \Omega, b, \delta_g$ then $\Omega, b \mid \delta_L, \delta_g$, using the sample of one block to condition the others. We also tested splitting the parameter space between $(\delta_L, \Omega)$ and $b$, based on the idea that sampling $b \mid \delta_L, \Omega, \delta_g$ is computationally cheap since it only involves evaluating the bias model and does not require simulating structure formation. However, even under the optimistic counting of zero model evaluation for a recomputation of the bias model, this approach performed worse than the other sampling methods.

\paragraph{MAMS, MCLMC.} We tested a variety of symplectic integrators\footnote{The specific interest of symplectic integrators for solving equations \ref{eqn:micro_langevin} is not direct since it is not formulated in canonical phase space $(\bs q,\bs p)$.} to solve the equations of motion~\ref{eqn:micro_langevin}. This includes the widely used KDK-Verlet integrator (2nd-order, 1 model evaluation per step), McLachlan integrator (2nd-order, 2 evaluations; see eq. 4.2 in \cite{mclachlan_1995_NumericalIntegrationOrdinary}), Blanes integrator (2nd-order, 3 evaluations; see eq. 44 in \cite{blanes_2014_IntegratorsHMC}), and Omelyan integrator (4th-order, 5 evaluations; see eq. 77 in \cite{omelyan_2003_SymplecticIntegrators}). The McLachlan and Omelyan schemes are known to achieve minimum norm within their respective orders \citep{takaishi_2006_IntegratorsHMCLQCD}, while the Blanes scheme offers best stability among 3-evaluation methods. 
In our tests, the McLachlan integrator consistently achieved the best performance, albeit by a modest margin, and is therefore adopted in the remainder of this work. As expected, more stable or higher-order integrators allowed for larger step sizes, but the gain was insufficient to offset their increased number of model evaluations. Additional results on the scaling of microcanonical samplers and the impact of integrators on sampling efficiency can be found in appendix \ref{sec:comp_mams_mn24}.

\paragraph{MAMS.} The step size $\eps$ is tuned to target a $65\%$ acceptance rate, theoretically optimal for isotropic Gaussian densities and corroborated in our tests.\footnote{\cite{robnik_2025_MetropolisAdjustedMicrocanonical} report higher values working well in their tests, which may be better suited for more complex densities.} The average trajectory length is set to heuristic $L= 0.3{\frac{N_\text{step}}{N_\text{eff}[\mu]}}\eps$ (with $N_\text{step} = N_\text{eval}/2$ for McLachlan integrator) \cite{robnik_2025_MetropolisAdjustedMicrocanonical}, where $N_\text{eff}[\mu]$ is a rough estimate of the (multivariate) ESS. 

\paragraph{MCLMC.} The step size $\eps$ is tuned to obtain an EEVPD below $10^{-6}$, ensuring negligible bias in the posterior distribution. A detailed discussion on the followed strategy to control the potential bias of MCLMC is provided in appendix~\ref{sec:bias_mclmc}. We set the momentum decoherence length according to a worst-case heuristic $L = 0.4 \frac{N_\text{step}}{\min_i N_\text{eff}[\mu_i]}\eps$, where $N_\text{eff}[\mu_i]$ is a rough estimate of the ESS of the $i$-th sampled variable. Contrary to adjusted samplers that yield one sample after a certain number of integration steps $L / \eps$, MCLMC yields one sample at each step. However, saving all these correlated samples is memory inefficient, so we apply chain thinning, retaining only every 16th step. We verify that this does not impact the performance metrics.

\begin{figure}[htbp]
    \centering
    \includegraphics[width=0.8\textwidth]{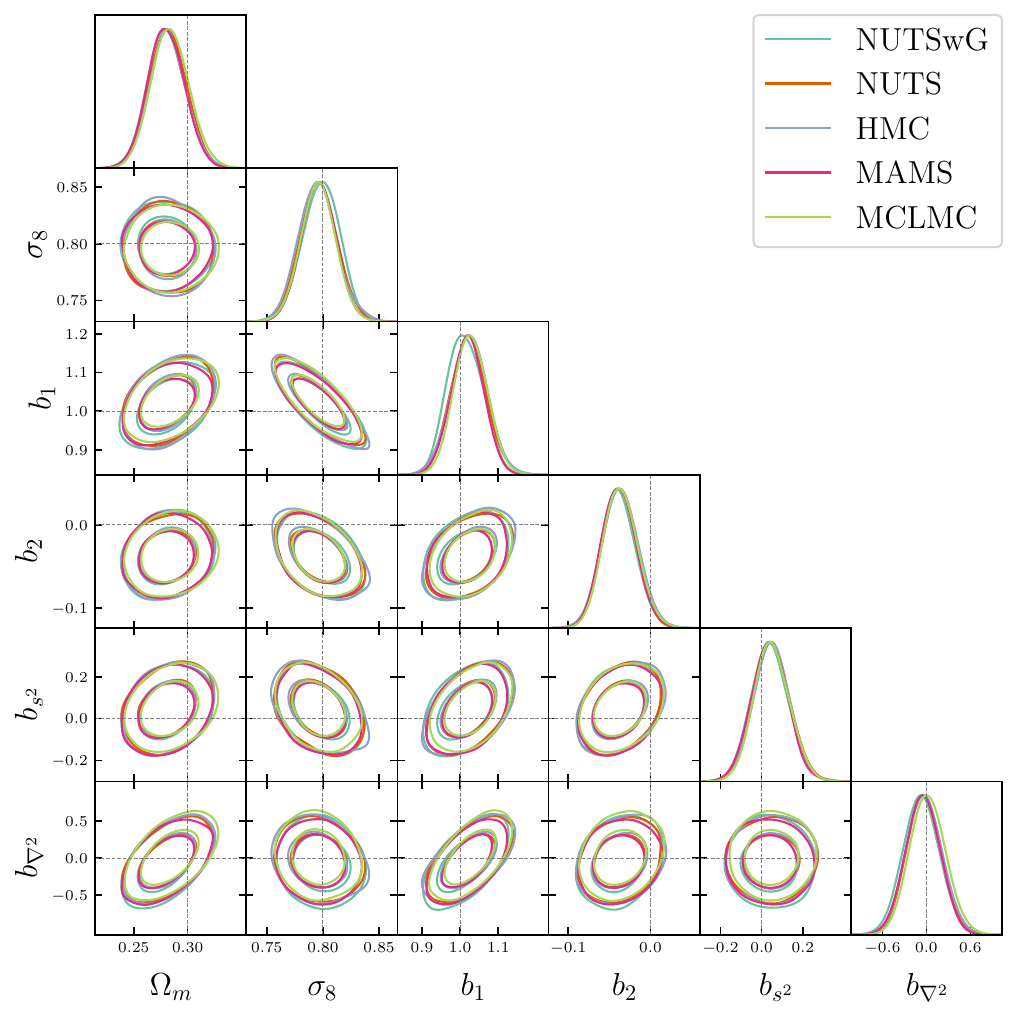}
    \caption{$68\%$ and $95\%$-credible regions of the cosmological and bias parameters posterior $\Omega, b \mid \delta_g$, obtained with the five considered samplers (NUTSwG in turquoise, NUTS in orange, HMC in indigo, MAMS in magenta, MCLMC in lime). The dashed lines denote the parameter true values used to generate the (synthetic) observation.}
    \label{fig:samplers_triangle}
\end{figure}

\begin{figure}[htbp]
    \centering
        \includegraphics[width=0.7\textwidth]{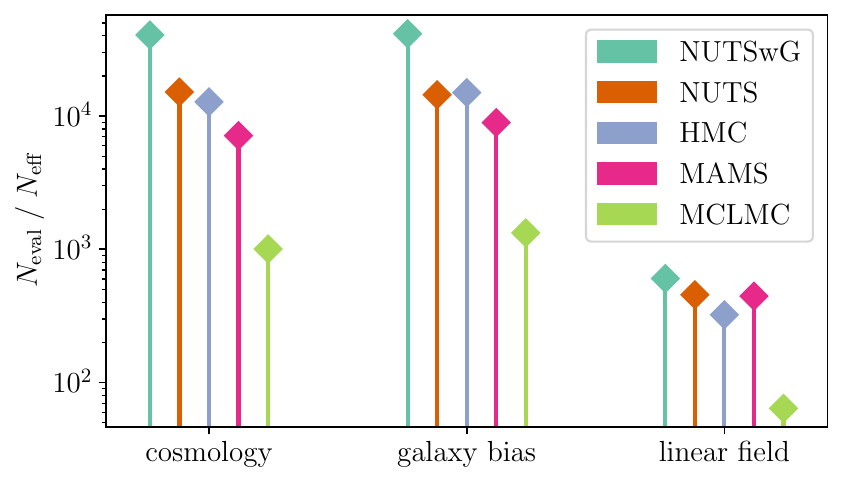}
        \caption{Number of model evaluations per effective sample for each sampler and each set of parameters (cosmology, galaxy bias and linear field).}
        \label{fig:samplers_ess}
\end{figure}

\medskip

The posterior $\Omega, b \mid \delta_g$ obtained for different samplers are in good agreement, as illustrated in figure~\ref{fig:samplers_triangle}. Moreover, figure~\ref{fig:samplers_ess} shows the number of model evaluations per effective sample $N_\text{eval} / N_\text{eff}$ for each sampler and each set of parameters. Both microcanonical samplers, MAMS and MCLMC, outperform their canonical counterparts. Notably, MCLMC surpasses all adjusted samplers, achieving a tenfold speedup in this case. In particular, when compared to MAMS, this suggests that for such high-dimensional settings, the efficiency gain stems primarily from the unadjusted nature of MCLMC, with its microcanonical nature playing a secondary role. 

Also, NUTSwG performs worse than standard NUTS, which is expected since MCMC within Gibbs sampling is theoretically suboptimal, particularly for correlated densities. While preconditioning and mass matrix tuning help mitigate correlations (see section~\ref{sec:expe_conds}), residual correlations may still contribute to this performance gap. Finally, HMC performs slightly better than NUTS, but this comparison benefits from the fact that HMC's trajectory length was tuned using prior NUTS runs. In practice, NUTS is generally preferable to HMC, as it eliminates the need for manual tuning of this sensitive parameter.

\subsection{Conditioning strategies}
\label{sec:expe_conds}

From this point onward, we focus on the MCLMC sampler, using it to benchmark four preconditioning strategies outlined in section~\ref{sec:intro_conds}, both with and without diagonal mass matrix tuning. As in the previous section, we verify that all resulting posteriors are consistent before reporting performance metrics.

\begin{figure}[htbp]
    \centering
    \includegraphics[width=0.7\textwidth]{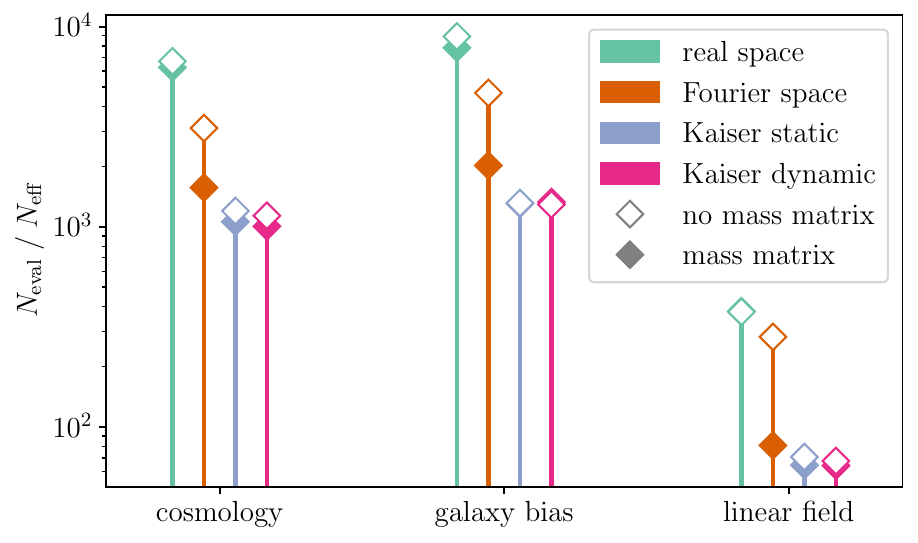}
    \caption{Number of model evaluations per effective sample for each set of parameters and each conditioning strategy: prior-conditioning in real space, prior-conditioning in Fourier space, static and dynamic Kaiser conditioning, all with (filled markers) and without (empty markers) mass matrix tuning.}
    \label{fig:conds_ess}
\end{figure}

\begin{table}[htbp]
    \centering
    \begin{tabular}{lrrr}
        \hline
        Strategy & $N_\text{eval} / N_\text{eff} \downarrow$ & $N_\text{eval} \times \operatorname{MSE}[\mu] \downarrow$ &  $N_\text{eval} \times \operatorname{MSE}[\sigma] \downarrow$\\
        \hline
        Best & 1005 \textcolor{white}{$\times 1.0$} & 879 \textcolor{white}{$\times 1.0$} & 390 \textcolor{white}{$\times 1.0$}\\
        w/o mass matrix & 1136 \textcolor{red}{$\times 1.1$} & 883 \textcolor{red}{$\times 1.0$} & 518 \textcolor{red}{$\times 1.3$}\\
        w/o dynamic & 1200 \textcolor{red}{$\times 1.1$} & 2256 \textcolor{red}{$\times 2.6$} & 579 \textcolor{red}{$\times 1.1$}\\
        w/o Kaiser & 3112 \textcolor{red}{$\times 2.6$} & 3314 \textcolor{red}{$\times 1.5$} & 1395 \textcolor{red}{$\times 2.4$}\\
        w/o Fourier & 6714 \textcolor{red}{$\times 2.2$} & 5504 \textcolor{red}{$\times 3.5$} & 1391 \textcolor{red}{$\times 1.0$}\\
        \hline
    \end{tabular}
    \caption{Cumulative ablation of the different conditioning strategies from our best result with dynamic Kaiser preconditioning and mass matrix tuning. The three metrics $N_\text{eval} / N_\text{eff}$, $N_\text{eval} \times \operatorname{MSE}[\mu]$ and $N_\text{eval} \times \operatorname{MSE}[\sigma]$ are computed for the cosmological parameters, and the true value used for computing MSEs are obtained from a long NUTS sampling run. Removing conditioning strategies consistently and cumulatively degrades performance.}
    \label{tab:ablation}
\end{table}

Figure~\ref{fig:conds_ess} presents the number of model evaluations per effective sample for each conditioning strategy and set of parameters. As expected, sampling in real space performs the worse, since voxels can be highly correlated. This strong correlation also limits the potential efficiency gains from diagonal mass matrix tuning, as reflected in the small improvement observed in that case.

In contrast, sampling in Fourier space is expected to largely decorrelate the sampled $\delta_L$ parameters. Although naive prior-conditioning in Fourier space can still lead to poor conditioning, since different scales are not equally constrained, it enables significant improvement when combined with diagonal mass matrix tuning. Even better performance is achieved through preconditioning in Fourier space using a static Kaiser model, which appears to capture nearly all the diagonal conditioning benefits of Fourier space, as indicated by the minimal additional gain from mass matrix tuning. Finally, preconditioning with a dynamic Kaiser model performs only slightly better, likely because cosmology and bias parameters are sufficiently constrained in our setup, making the conditioning of the linear field weakly sensitive to their variations. For MCLMC in particular, better conditioning correlates with greater stability in numerical integration, allowing larger step sizes for the same nominal EEVPD, thereby enhancing sampling efficiency.

All metrics for the cosmological parameters are reported in table~\ref{tab:ablation}. The ground-truth values used for computing MSEs are obtained from a long NUTS sampling run, which is asymptotically unbiased. The metrics are consistent, with both $N_\mathrm{eval} / N_\mathrm{eff}$ and $N_\mathrm{eval} \times \operatorname{MSE}[\mu]$ values of the same order, indicating no detectable bias in MCLMC sampling. Ultimately, we demonstrate that effective conditioning can improve sampling efficiency by approximately a factor of ten.

\subsection{Scaling effects}
\label{sec:expe_scaling}

Now that we have more insights into the performance of MCMC samplers and conditioning strategies for field-level inference, we aim to fully assess their applicability to upcoming galaxy surveys. In particular, we are interested in the scalability of MCLMC sampling with respect to dimensionality and modeling complexity.

\begin{figure}[htbp]
    \centering
    \includegraphics[width=0.99\textwidth]{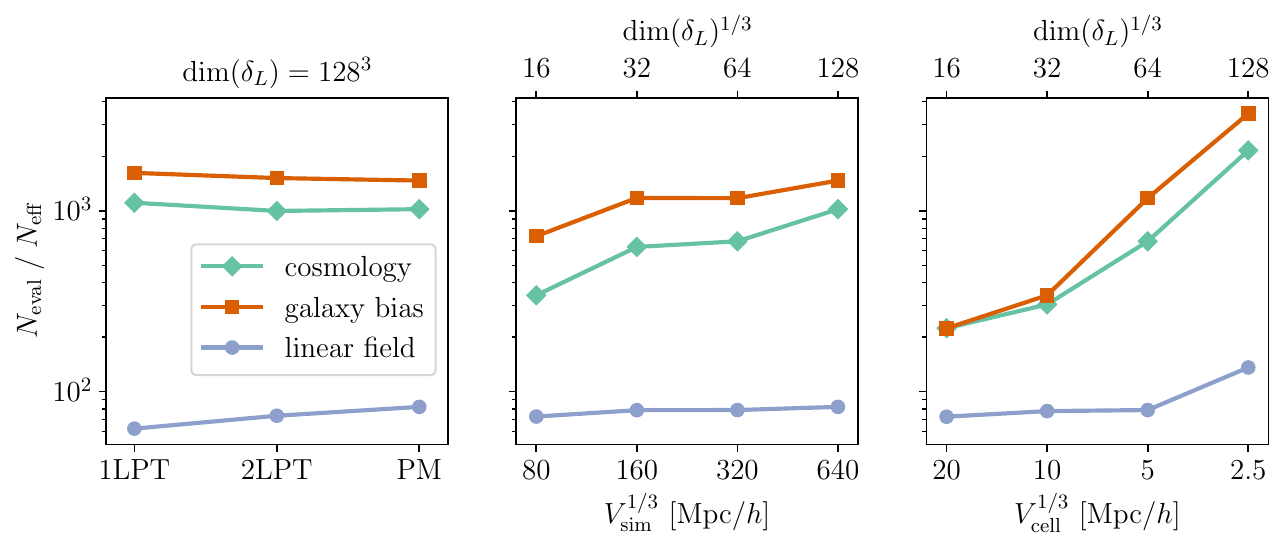}
    \caption{Number of model evaluations per effective sample for different model settings. Left: scaling with the complexity of the formation model at fixed resolution of $\qty{5}{\mega\pc/\h}$. Middle: scaling with the simulated volume for PM model. Right: scaling with the voxel resolution for PM model.}
    \label{fig:scaling_ess}
\end{figure}

Figure~\ref{fig:scaling_ess} (left panel) shows the number of model evaluations per effective sample for various LSS formation models (1LPT, 2LPT and PM) at a fixed volume of $V_\mathrm{sim}=(\qty{640}{\mega\pc/\h})^3$ and voxel resolution of $\qty{5}{\mega\pc/\h}$ (hence $\operatorname{dim}(\delta_L) = 128^3$). We observe that in this setting, the sampling efficiency is largely independent of the complexity of the LSS formation model. Notably, we recover a similar magnitude in sampling efficiency as presented in~\cite{Bayer_Seljak_Modi_2023} (below $10^2$ and $10^3\, N_\mathrm{eval} / N_\mathrm{eff}$ for linear field and cosmology respectively). One may notice a slight increase in the sampling cost of the linear field, which could be attributed to the increased non-linearity of the forward model for LSS formation.

Figure~\ref{fig:scaling_ess} (middle panel) shows the same metric for different volumes, at a fixed voxel resolution of $\qty{5}{\mega\pc/\h}$ with the PM formation model. We observe almost no increase in the sampling cost of the linear field with respect to dimensionality. Since increasing the volume at fixed resolution primarily adds independent wavevectors to infer from, this corroborates the theoretical expectation that MCLMC efficiency remains independent of dimension for product densities. However, for cosmological and bias parameters, we note a slight increase in sampling cost with dimensionality, approximately scaling as $d^{1/5}$. This is possibly related to the growing constraining power on these parameters, which reveals non-trivial correlations.

This effect is more pronounced when increasing resolution instead of volume. Figure~\ref{fig:scaling_ess} (right panel) shows sampling efficiency at a fixed volume of $V_\mathrm{sim} = (\qty{320}{\mega\pc/\h})^3$ and varying resolutions. We observe a clear increase in sampling cost with dimensionality, approximately scaling as $d$ for cosmological and bias parameters. This indicates that field-level inference can become intrinsically more challenging at non-linear scales.

We must note that all these results are obtained for a mean galaxy density of $\bar{n}_g = \qty{5e-4}{(\mega\pc/\h)^3}$, which is half the value used in previous sections and therefore yields twice the observational variance. This choice was made because, with the previously considered density of $\bar{n}_g = \qty{e-3}{(\mega\pc/\h)^{-3}}$ we found that for large volumes $V_\mathrm{sim} \geq (\qty{640}{\mega\pc/\h})^3$ or high resolution $V_\mathrm{cell} \leq (\qty{2.5}{\mega\pc/\h})^3$ and for the PM model only, the posterior exhibited some multimodality, with the chains getting stuck in different local modes\footnote{We further checked this with a NUTS sampling run in case this was due to a MCLMC bias.} (local in the sense that some modes admit visibly lower density values than others). In practice, this problem could be addressed in several ways: (i) adopting an annealing or tempering strategy (as performed in \citep{bayer_2023_AnnealingVelocityDensityReconstruction, Porqueres_2020_borg_lyman}) to help the chains converge to or navigate between global modes; (ii) smoothing or noising the observations (at the cost of reducing constraining power). We think the first option would make fair comparisons more difficult in this benchmark, and that it is worth being benchmarked on its own. Therefore, we opted for introducing more observational noise (remaining within the typical galaxy density range of current spectroscopic surveys) to focus on the general scalability aspects of the problem.

On an indicative basis, with our best sampler and conditioning strategy, in the most computational intensive setting ($\operatorname{dim}(\delta_L) = 128^3$ with 5-step PM displacement), a complete posterior can be obtained in less than 4 hours on a 8$\times 80 $GB GPU node. For the same setting and conditioning strategy, NUTS took more than 80 hours.

\section{Conclusion}
\label{sec:conclusion}

We have benchmarked various state-of-the-art samplers for field-level cosmological inference from galaxy redshift surveys, emphasizing the critical role of preconditioning latent variables, particularly the initial density field, $\delta_L$. We find that Fourier-space sampling of $\delta_L$, combined with a linear rescaling which assumes a simple Kaiser model, improves sampling efficiency by an order-of-magnitude compared to naive real-space sampling. In this setup, tuning the mass matrix provides only marginal additional gains. However, we note that for real survey applications where observational noise varies across the footprint, this preconditioning may need refinement, and mass matrix tuning could become more relevant.  

Among the samplers tested (HMC, NUTS, NUTS within Gibbs, MAMS, and MCLMC), the recently proposed MCLMC achieves the highest efficiency, outperforming other methods by approximately an order-of-magnitude. This gain mostly arises from MCLMC being run unadjusted, at the cost of introducing a potential bias in the cosmological posterior. However, this bias can be effectively controlled and reduced to negligible levels with a modest number of additional sampling steps. For a relevant modeling at reasonable scales ($\geq \qty{5}{\mega\pc/\h}$ with a PM scheme and 2nd order Lagrangian bias), MCLMC typically requires \(\approx 10^3\) model evaluations per effective sample, mildly dependent of problem dimensionality, making it highly scalable for real-data applications. Nonetheless, we see that extending the inference to smaller physical scales may degrade efficiency, potentially requiring the use of tempering strategies to maintain robust performance. Similar complication may arise when considering more accurate small-scale modeling.

The benchmark code is publicly available\footnote{\hurl{github.com/hsimonfroy/benchmark-field-level}} and is intended to serve as a basis for future comparisons of alternative sampling methods. The forward model will be regularly augmented, in line with \texttt{JaxPM} developments, including the forthcoming implementation of massively parallel FFT calculations\footnote{\hurl{github.com/DifferentiableUniverseInitiative/jaxDecomp}} and improvements in the redshift-space galaxy density modeling. 

Applying these methods to real survey data will indeed require additional developments, including a proper light-cone model, refined galaxy-matter connection at the relevant scales in redshift space, and the incorporation of survey selection and observational systematics. These enhancements will be the focus of future work.

\acknowledgments
The authors thank Marco Bonici and Natalia Porqueres for thoughtful discussions on field-level inference, Jakob Robnik for his feedback on the benchmark, and Wassim Kabalan for valuable insights on numerical implementation with JAX. They are especially grateful to Adrian Bayer for his constructive feedback and helpful comments on the manuscript. This work was granted access to the HPC resources of IDRIS under the allocations 2024-AD010414933 and 2025-475 AD010414933R1 made by GENCI. 

\appendix

\section{Additional comparisons}
\label{sec:add_comp}

\subsection{Non-trivial selection}
\label{sec:comp_selec}

In this work, we benchmarked field-level inference in an idealized periodic box setting.
selection and flat-sky setting. However, realistic surveys have significant masking and variable depth, which break translation invariance and introduce long-range coupling of wavevector amplitudes. 
Both our MCMC initialization and preconditioning rely on a linear diagonal approximation in Fourier space of our model given by \ref{eqn:kaiser_post}, that is challenged in such more realistic setting. To prospect further developments, we briefly explore in this section the impact of curved-sky geometry and non-trivial selection functions on sampling efficiency.

\begin{figure}[htbp]
    \centering
    \includegraphics[width=\textwidth]{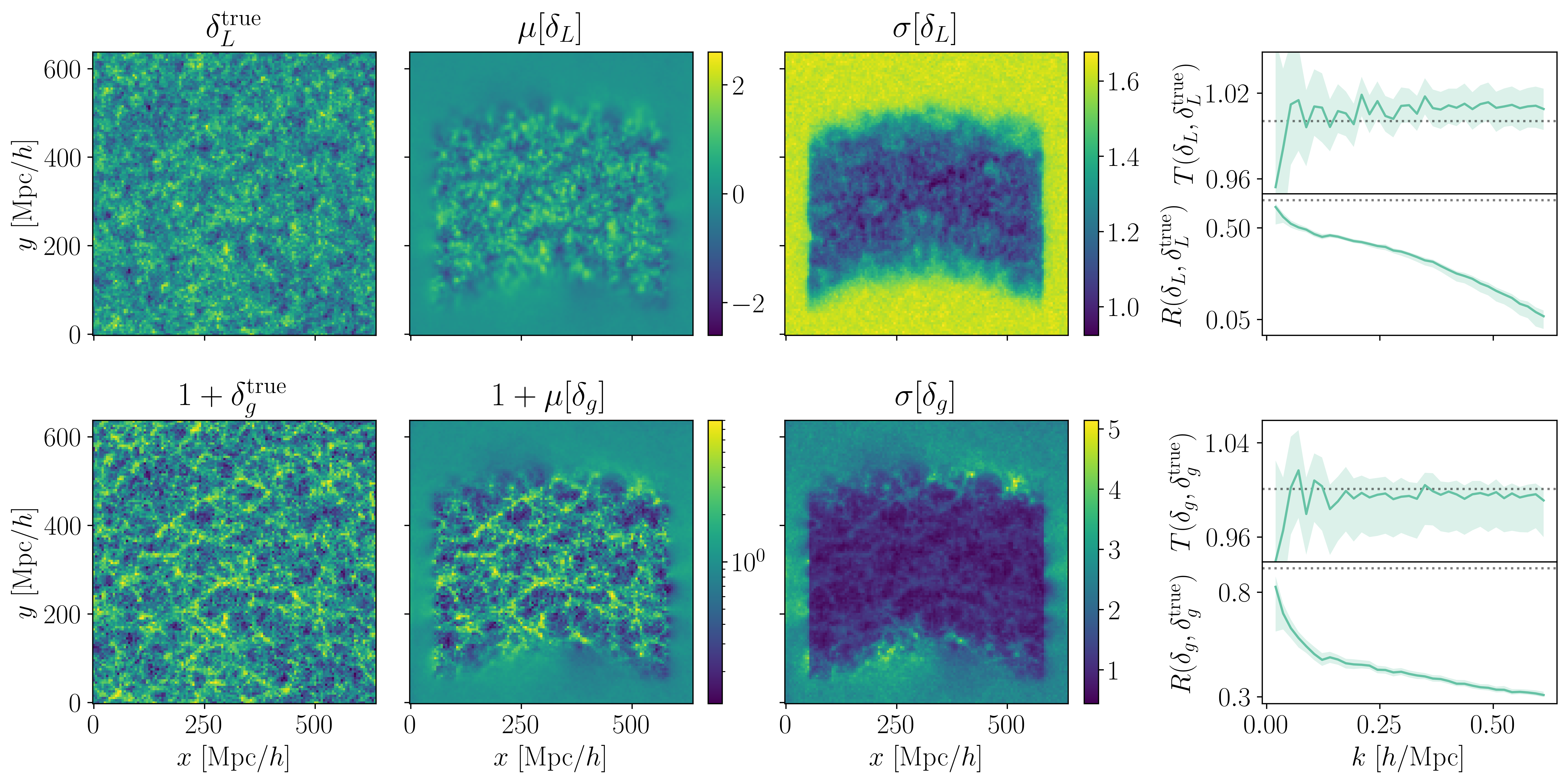} 
    \caption{Posterior of (top row) the initial linear matter field $\delta_L$ and (bottom row) the evolved galaxy density field $\delta_g$, obtained from field-level inference as in figure \ref{fig:post_fields}, but for a non-trivial selection function. Both the linear matter and galaxy density fields are well reconstructed on large scales within the selection (lower standard deviation, transfer function and spectra correlation close to 1), while they are less constrained outside of the selection (higher standard deviation).}
    \label{fig:post_fields_selec}
\end{figure}

We take a selection window $W$ (normalized to unit mean value over voxels, $\bar W=1$) consisting in a shell cut around the observer, and RSD are applied along the varying line-of-sight. Out model outputs an observable galaxy field $N_g$ (in number of galaxies per voxel), from which we can define the fluctuation field $F_g := N_g - \bar N_g W$, such that equation \ref{eqn:kaiser_post} becomes
\begin{align}
    &\underbrace{\cN (F_g \mid \bar N_g W B_K \delta_\mathrm{L}, \bar N_g W )}_{\text{likelihood}}\;\underbrace{\cC \cN(\delta_\mathrm{L} \mid 0, P_L V_\mathrm{cell}^{-1})}_{\text{prior}}\nonumber \\= &\underbrace{\cC\cN(\delta_\mathrm{L} \mid \mu, \Sigma)}_{\text{posterior}}\; \underbrace{\cN(F_g \mid 0, \bar N_g W + \bar N_g^2 V_\mathrm{cell}^{-1} W  B_K P_L B_K W)}_{\text{evidence}}
\label{eqn:kaiser_post_selec}
\end{align}
with\begin{align*}
\Sigma&:= (\bar N_g B_K W B_K + P_L^{-1} V_\mathrm{cell})^{-1} \\
\mu &:= \Sigma B_K F_g
\end{align*}
The main computational obstacle in applying this formula is that $\Sigma$ is non-diagonal in both real and Fourier space, because $W$ is diagonal in real space, whereas $P_L$ is diagonal in Fourier space. To alleviate this, we approximate $W$ by the constant $\tilde W := \bar{W^2}^{1/2} = (1+\sigma_W^2)^{1/2} \geq 1$, the root-mean-square of the window over voxels. Also, in curved-sky, $B_K := (b_E + (\hat{\bs k} \cdot \hat{\bs \eta})^2 f) D$ is neither diagonal in real space nor in Fourier space, but we can make the approximation that the line-of-sight $\hat{\bs \eta}$ is fixed to the line-of-sight towards the center of the survey volume, so that we keep $B_K$ diagonal in Fourier space. The rest of the implementation is left unchanged.

Figure \ref{fig:post_fields_selec} illustrates field-level inference for this more realistic setting. As we expect, the effect of the selection is clearly seen in the posterior standard deviation of the initial field and the galaxy density field: regions within the selection are better constrained than regions outside, which are more prior-dominated. More generally, the constraints are broader on all the metrics when compared to the no selection case in figure \ref{fig:post_fields} (larger standard deviation at pixel level, larger variability in transfer function, faster decay of spectra correlation), which is consistent with a loss of information due to selection.

Finally, figure \ref{fig:selection_scaling} shows the number of model evaluations per effective sample with respect to the dimension and simulated volume, for a PM model with and without selection and a fixed resolution of $\qty{5}{\mega\pc/\h}$. We observe a similar scaling between selections, but the non-trivial selection seems to degrade a bit the efficiency. This indicates that, while field-level inference may be more computationally demanding in realistic survey settings, the overall scaling of sampling performance with dimensionality remains largely unaffected.

\subsection{Scaling of adjusted and unadjusted samplers}
\label{sec:comp_mams_mn24}

\begin{figure}[htbp]
     \centering
     \begin{subfigure}[t]{0.48\textwidth}
         \centering
         \includegraphics[width=\textwidth]{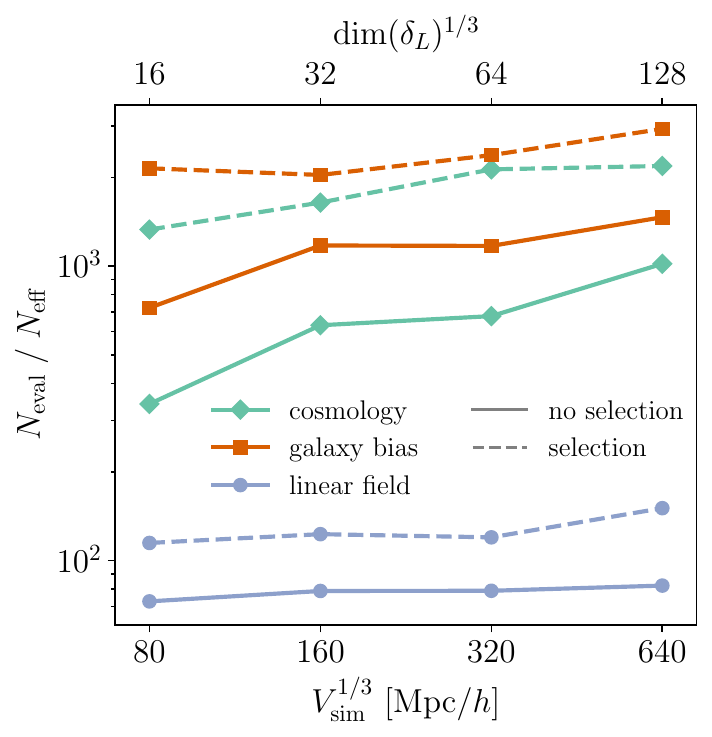}
         \caption{Scaling with the dimension and simulated volume, for a PM model with and without survey selection, at fixed resolution of $\qty{5}{\mega\pc/\h}$.}
         \label{fig:selection_scaling}
     \end{subfigure}
     \hfill
     \begin{subfigure}[t]{0.48\textwidth}
         \centering
         \includegraphics[width=\textwidth]{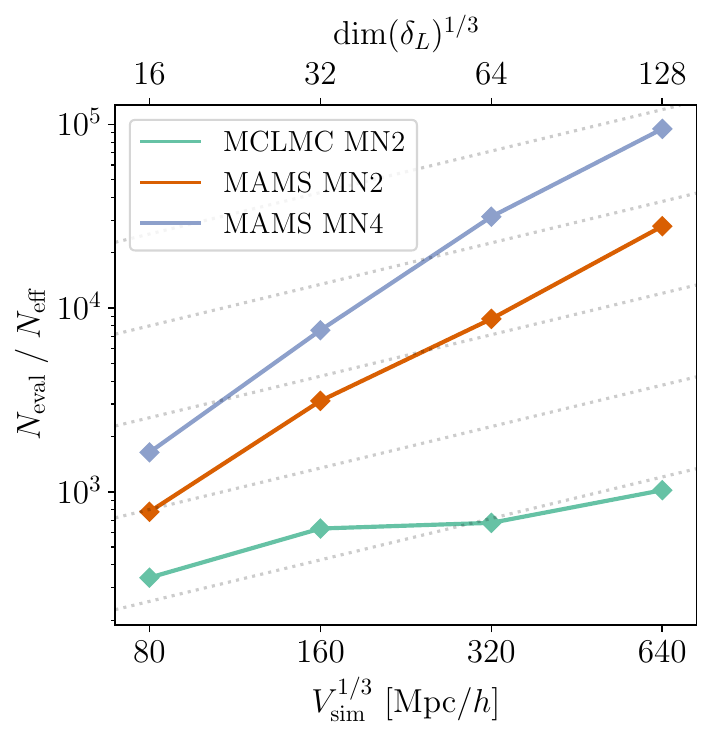}
         \caption{Scaling for MCLMC and MAMS with McLachlan (MN2) and Omelyan (MN4) integrators. Dotted lines represent the $d^{1/4}$ scaling, expected for MAMS and HMC with 2nd order integrators.}
         \label{fig:scaling_mamas_mclmc}
     \end{subfigure}
     \caption{Number of model evaluations per effective sample with respect to dimension and simulated volume, for different models and MCMC sampler settings.}
\end{figure}

To illustrate the relevance of unadjusted sampling for field-level inference, we investigate whether higher-order integrators could help adjusted samplers to achieve efficiencies comparable to unadjusted ones. For this, we compare the unadjusted MCLMC to the adjusted MAMS, with McLachlan (alias 2nd-order Minium Norm, MN2) or Omelyan (4th-order Minimum Norm, MN4) integrators. The step size $\eps$ is tuned to target 65\% and 80\% acceptance rates for MN2 and MN4 respectively, which are theoretically optimal for isotropic Gaussian densities \citep{robnik_2025_MetropolisAdjustedMicrocanonical}. The average trajectory length is set following the heuristic $L \propto \frac{N_\text{step}}{N_\text{eff}}\eps$, as mentioned in \ref{sec:expe_samplers}. We also tested a simpler $L \propto \eps$ heuristic which happened to be worse for both integrators.

Figure \ref{fig:scaling_mamas_mclmc} shows the number of model evaluations per effective sample with respect to the simulated volume, for a PM model (without survey selection) at fixed resolution of $\qty{5}{\mega\pc/\h}$, and for MCLMC and MAMS samplers with MN2 and MN4 integrators.
As expected, unadjusted sampling exhibits strikingly better scaling with dimensionality than adjusted one. Also, both MCLMC and MAMS scale a bit worse than their respective theoretical rates for independent dimensions, respectively constant and $d^{1/4}$, which suggests potential non-trivial correlations appearing with growing constraining power. Moreover, MN4 integrator seems to perform consistently worse than MN2 integrator. This suggests that while higher-order integrators can be theoretically appealing by allowing for larger step sizes, they can be numerically less stable and may not provide sufficient gain to offset their increased number of evaluations.

\section{Hamiltonian formulation of the isokinetic Langevin dynamics}
\label{sec:gik_hamilt}

Here we show that we can rederive from an Hamiltonian the MCLMC equations of motion \citep{robnik_2024_FluctuationDissipationMicrocanonical}, without resorting to a momentum-dependent time rescaling used in \citep{steeg_2021_HamiltonianDynamicsNonNewtonian, robnik_2022_MicrocanonicalHamiltonianMonte}. We follow the idea of \citep{morriss_1998_ThermostatsAnalysisApplication} and consider a particle of position $\bs q$, momentum $\bs p$, and mass matrix $M$, evolving in a $d$-variate potential $U$, with the Gaussian IsoKinetic (GIK) Hamiltonian

\begin{equation}
    \cH(\bs q, \bs p) = \frac {\bs p\T M^{-1} \bs p} {2 m(\bs q)} - K m(\bs q)
\end{equation}
where $m := e^{-U}$ and $K$ is a constant. Note that the adjective \emph{Gaussian} in GIK refers to Gauss's principle of least constraint, rather than to the Gaussian distribution. 

Setting the auxiliary momentum $\bs u := m^{-1} M^{-1/2} \bs p$ and initial energy $E:= \cH(\bs q_0, \bs p_0)$, we obtain from energy conservation that the kinetic energy is $\dot{\bs q}\T M \dot{\bs q} / 2= |\bs u|^2 / 2 = K + E /m(\bs q)$. For the canonical phase state $(\bs q, \bs p)$, Hamilton equations are then
\begin{equation}
\begin{cases}
\dot{\bs q} = \partial_{\bs p}\cH = (m(\bs q) M)^{-1} \bs p\\
\dot{\bs p} = - \partial_{\bs q}\cH = - \nabla U(\bs q)\,  m(\bs q)\, (2K + E / m(\bs q))
\end{cases}
\end{equation}
which translates for state $(\bs q, \bs u)$ into
\begin{equation}
\begin{cases}
\dot{\bs q} = M^{-1/2} \bs u\\
\dot{\bs u} = - ((2K + E/ m(\bs q))I - \bs u \bs u\T) M^{-1/2} \nabla U(\bs q)
\end{cases}
\end{equation}

This dynamics trivially admits the {microcanonical ensemble} $\pp_\mathrm{MC}(\bs q, \bs p) \propto \delta(\cH(\bs q,\bs p) - E)$ as a stationary density. Moreover, when starting at kinetic energy $|\bs u_0|^2 / 2 = K$, i.e. $E = 0$, this ensemble coincides with the {isokinetic ensemble} \citep{morriss_1998_ThermostatsAnalysisApplication}
\begin{equation}
    \pp_\mathrm{MC}(\bs q, \bs p) \propto \delta(\cH(\bs q,\bs p)) \propto \pp_\mathrm{IK}(\bs q, \bs u) \propto  e^{-(d-1) U(\bs q)} \delta(|\bs u|^2/2 - K)
\end{equation}
In particular, if ergodicity holds, the particle position $\bs q$ samples the potential $U$ at the effective temperature $(d-1)^{-1}$. Therefore, by replacing $U$ by $U/(d-1)$, setting $K=1/2$, and adding Gaussian white noise as in \citep{robnik_2024_FluctuationDissipationMicrocanonical}, we recover the desired isokinetic Langevin equations \ref{eqn:micro_langevin} that sample from $\pp_\mathrm{IK}(\bs q, \bs u) \propto e^{-U(\bs q)} \delta(|\bs u|^2 - 1)$.

\section{Handling bias in unadjusted Langevin dynamics}
\label{sec:bias_mclmc}

In section~\ref{sec:expe_samplers}, we showed that unadjusted MCLMC can efficiently provide posterior samples without noticeable bias. Here, we illustrate the biased behavior of MCLMC with a poor choice of step size, or equivalently an inappropriate Energy Error Variance Per Dimension (EEVPD) threshold.

\begin{figure}[htbp]
    \centering
    \includegraphics[width=0.7\textwidth]{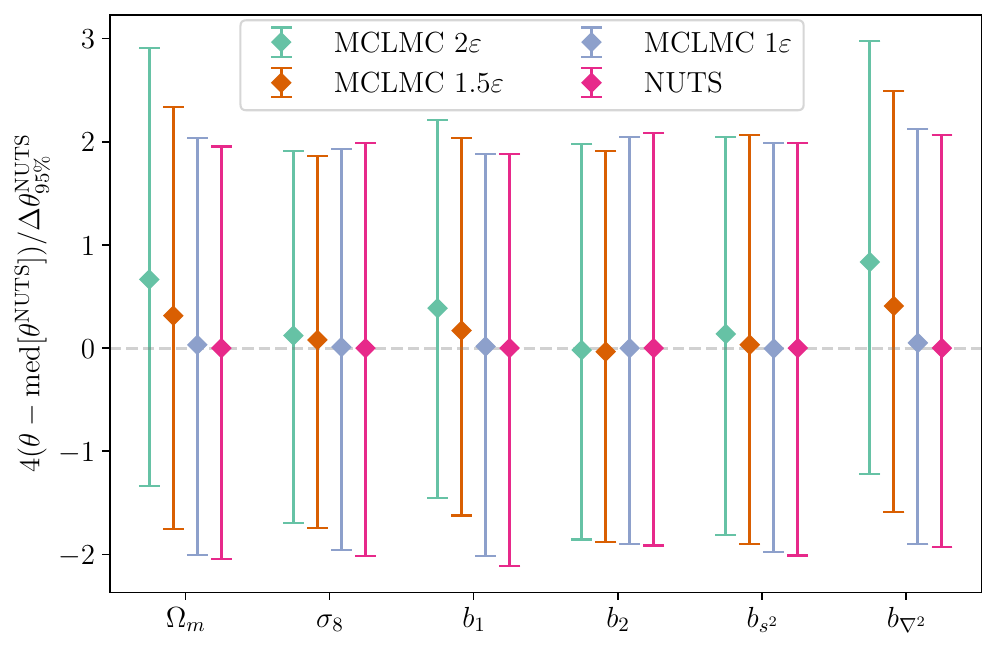}
    \caption{Comparison of posterior median and $95\%$-credible interval with respect to NUTS for the cosmological and bias parameters $\Omega, b \mid \delta_g$. We consider different step sizes: $1\eps, 1.5\eps, 2\eps$ given $\eps = 33$ the nominal step size. The corresponding EEVPD are $10^{-6}, 4\times 10^{-6}, 10^{-5}$.}
    \label{fig:bias_mclmc}
\end{figure}

Figure~\ref{fig:bias_mclmc} compares the posterior distributions of cosmological and bias parameters obtained from MCLMC with different step sizes to those obtained from NUTS, which is adjusted and therefore asymptotically unbiased. The model and sampler configurations are identical to those in section~\ref{sec:expe_samplers}.  We observe that our chosen EEVPD threshold of $\braket{\Delta E^2} /d = 10^{-6}$ produces median and credible interval estimates nearly identical to those from NUTS, with differences likely dominated by Monte Carlo error. However, when using step sizes that are $1.5\times$ or $2\times$ larger, a clear bias appears in the estimation of some parameters, particularly $\Omega_\mathrm{m}$, $b_1$, and $b_{\nabla^2}$.

We showed in section~\ref{sec:expe_samplers} that for high-dimensional applications, NUTS is a magnitude less efficient than MCLMC, making it unsuitable for testing potential MCLMC bias. As a best practice, we recommend performing MCLMC sampling with at least two different EEVPD values, e.g. $\braket{\Delta E^2} /d$ and $\braket{\Delta E^2} /(10 d)$, and checking for convergence of the posterior for the parameters of interest. Based on our experiments, once we identified an appropriate EEVPD threshold for our field-level inference problem (including preconditioning), we did not need to tune it further across the different settings considered in section~\ref{sec:expe_scaling}.

\bibliography{biblio}{}
\bibliographystyle{JHEP}

\end{document}